\documentclass[letterpaper,twocolumn,10pt]{article}
\RequirePackage{fix-cm}
\usepackage{usenix-2020-09}
\microtypecontext{spacing=nonfrench}

\usepackage{graphicx} 
\DeclareGraphicsExtensions{.pdf,.eps,.png,.jpg}

\usepackage{amsmath, amssymb} 
\usepackage{booktabs} 
\usepackage[inline]{enumitem} 
\usepackage{balance} 
\usepackage{multirow} 
\usepackage{multicol} 
\usepackage{longtable}
\usepackage{csquotes}

\usepackage{siunitx}
\DeclareSIUnit[round-mode = places, round-precision = 0]{\percent}{\%}
\DeclareSIUnit[number-unit-product = \,]{\USD}{\textsc{USD}}
\DeclareSIUnit{\dollar}{\$}
\DeclareSIPrefix{\million}{\text{ million }}{6}
\DeclareSIPrefix{\billion}{\text{ billion }}{9}
\newcommand{\pct}[1]{\num[round-mode = places, round-precision = 0, round-minimum = 1]{#1}}

\usepackage{caption}
\usepackage{subcaption}
\hypersetup{%
  pdflang={en-US},
  pdfkeywords={privacy; user study; Google My Activity},
  pdfdisplaydoctitle=true,
  bookmarksnumbered,
  colorlinks = false,
  pdfborder = {0 0 0},
  pdfstartview={FitH},
  breaklinks=true,
  hypertexnames=false
}

\newcommand*{\appref}[1]{\hyperref[#1]{Appendix~\ref*{#1}}}

\newcommand{\ie}{i.\,e.}
\newcommand{\eg}{e.\,g.}

\newcommand{\etal}{et~al.\@\,}

\newsavebox{\equalrowbox}
\newcolumntype{E}[1]{>{\begin{lrbox}{\equalrowbox}}l<{\end{lrbox}\raisebox{0pt}[#1]{\usebox{\equalrowbox}}}}%
\newcolumntype{H}{>{\iffalse}c<{\fi}@{}}

\newlist{compactitem}{itemize}{5}
\setlist[compactitem]{leftmargin=*, nosep}
\setlist[compactitem, 1]{label=\textbullet}
\setlist[compactitem, 2]{label=\textendash}
\setlist[compactitem, 3]{label=\textasteriskcentered}
\setlist[compactitem, 4]{label=\textperiodcentered}

\makeatletter
\let\orgItem\item
\NewDocumentCommand\fixedItem{ o }{%
   \IfNoValueTF{#1}%
      {\orgItem}
      {\orgItem[#1]\def\@currentlabel{#1}}
}
\makeatother

\newlist{questions}{enumerate}{3}
\setlist[questions]{align=left, labelwidth=2em, labelsep=.5em, listparindent=0pt, itemindent=0pt, leftmargin=!, before=\let\item\fixedItem}
\setlist[questions, 1]{labelindent=0pt, label=\textbf{Q\arabic*}, widest=99}
\setlist[questions, 2]{labelindent=-2.5em, label*=\textbf{\_\Alph*}, widest=26}
\setlist[questions, 3]{labelindent=-2.5em, label*=\textbf{\_\roman*}, widest=9}

\newlist{answers}{itemize}{1}
\setlist[answers]{leftmargin=*, nosep, align=left, label=$\bigcirc$}
\newlist{answers*}{itemize*}{1}
\setlist[answers*]{label=$\bigcirc$}

\begin{document}
\day=28 \month=05 \year=2021
\date{}

\title{
    \textbf{\Large Are Privacy Dashboards Good for End Users? \\ Evaluating User Perceptions and Reactions to Google's My Activity \\ (Extended Version)}
}

\hypersetup{
pdfauthor={Florian M. Farke, David G. Balash, Maximilian Golla, Markus Duermuth, and Adam J. Aviv}
}

{
\renewcommand{\thefootnote}{\fnsymbol{footnote}}
\author{
  {\rm Florian M. Farke$^{\dagger}${\footnotemark[1]}, David G. Balash$^{\S}${\footnotemark[1]},
    Maximilian Golla$^{\ddagger}$, Markus D\"{u}rmuth$^{\dagger}$, Adam J. Aviv$^{\S}$}\\
  {\normalsize $^\dagger$ Ruhr University Bochum, $^{\S}$ The George Washington University, $^\ddagger$ Max Planck Institute for Security and Privacy}
} 

\maketitle
\footnotetext[1]{The first two authors contributed equally to the paper.}
}


\newcommand{\figstudydesign}[0]{
\begin{figure*}[t]
\centering
\includegraphics[width=0.8\linewidth]{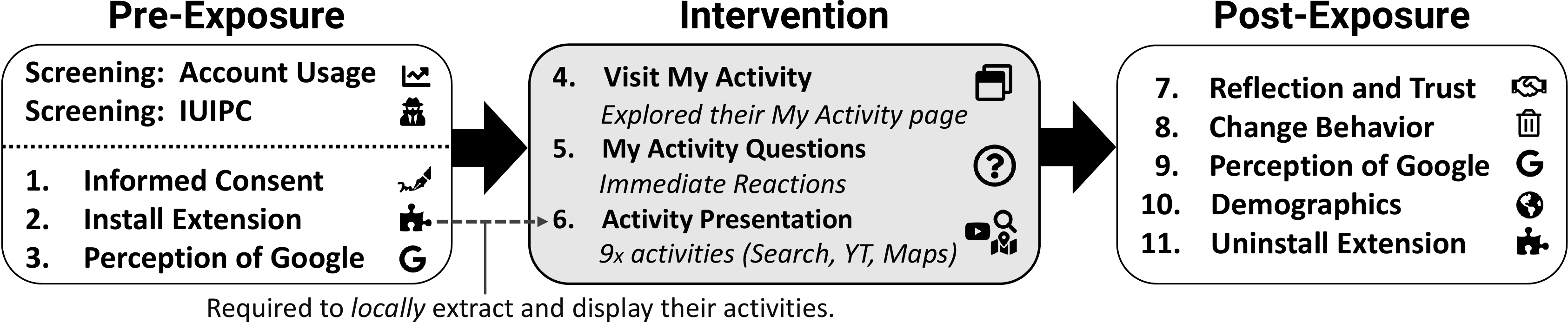}
\caption[Main study  protocol]{\label{fig:study-design}
Main Study: The study was divided into three parts.
During the intervention part, participants visited their own My Activity page and were questioned about nine of their activities (three per category) from Google Search, YouTube, and Maps.
}
\end{figure*}
}

\newcommand{\figbetterUnderstandingBar}[0]{
\begin{figure}[t]
\centering
\includegraphics[width=\columnwidth]{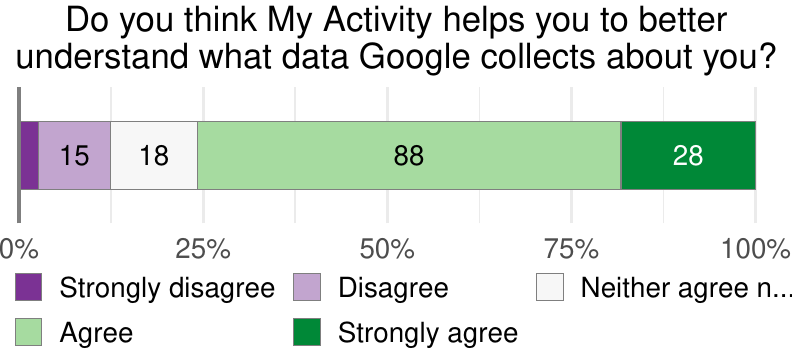}
\caption[Better understanding after visiting My Activity]{\label{fig:betterUnderstanding-bar}
Roughly \SI{75}{\percent} of the participants stated that My Activity helps them to better understand what data Google is collecting about them.
Only around \SI{12}{\percent} do not think My Activity aids their understanding.
}
\end{figure}
}

\newcommand{\figConcernAndBenefitBar}[0]{
\begin{figure*}[thbp]
\centering
\begin{subfigure}[t]{\columnwidth}
    \centering
    \includegraphics[width=\columnwidth]{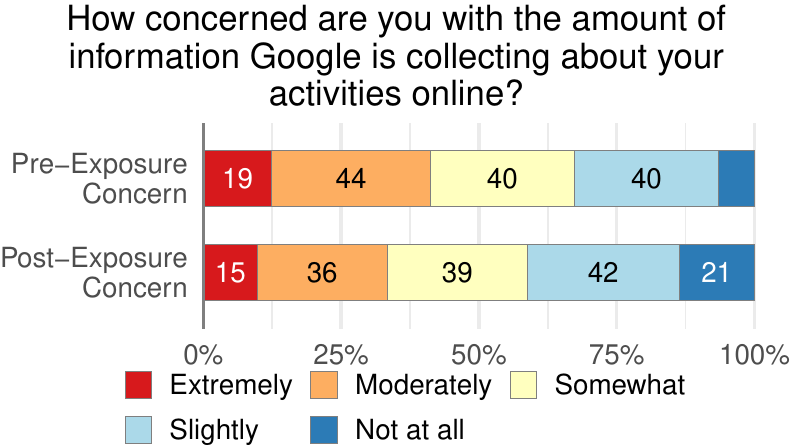}
    \caption[Level of concern before and after visiting My Activity]{\label{fig:collectingConcern-bar}
        Level of concern before and after visiting My Activity.
    }
\end{subfigure}
\hfill
\begin{subfigure}[t]{\columnwidth}
    \centering
    \includegraphics[width=\columnwidth]{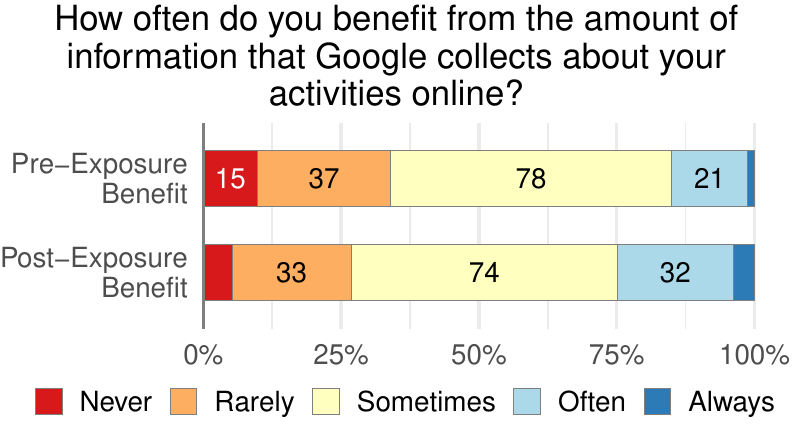}
    \caption[Frequency of benefit before and after visiting My Activity]{\label{fig:frequencyBenefit-bar}
        Frequency of benefit before and after visiting My Activity.
    }
\end{subfigure}
\caption[Level of concern before and after visiting My Activity]{\label{fig:concern-and-benefit}
Proportions of the participants' assessment of
\begin{enumerate*}[label=(\alph*)]
\item the level of concern (\ref{app:main-survey:Q2} \& \ref{app:main-survey:Q26}) and
\item the frequency of benefit (\ref{app:main-survey:Q3} \& \ref{app:main-survey:Q27})
\end{enumerate*}
before and after visiting the My Activity dashboard.
}
\end{figure*}
}

\newcommand{\figConcernAndBenefitAlluvium}[0]{
\begin{figure*}[thbp]
\centering
\begin{subfigure}[t]{\columnwidth}
    \includegraphics[width=\textwidth]{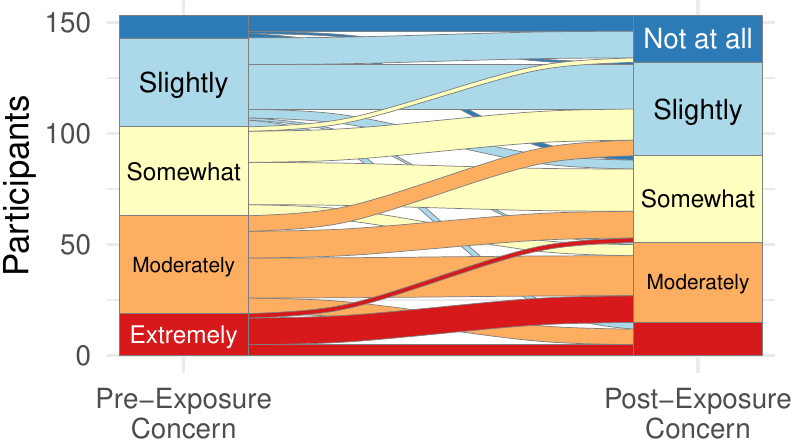}
    \caption[Perceived benefit alluvium plot]{\label{fig:collectingConcern-alluvium}
        Level of concern alluvium plot.
    }
\end{subfigure}
\hfill
\begin{subfigure}[t]{\columnwidth}
    \includegraphics[width=\textwidth]{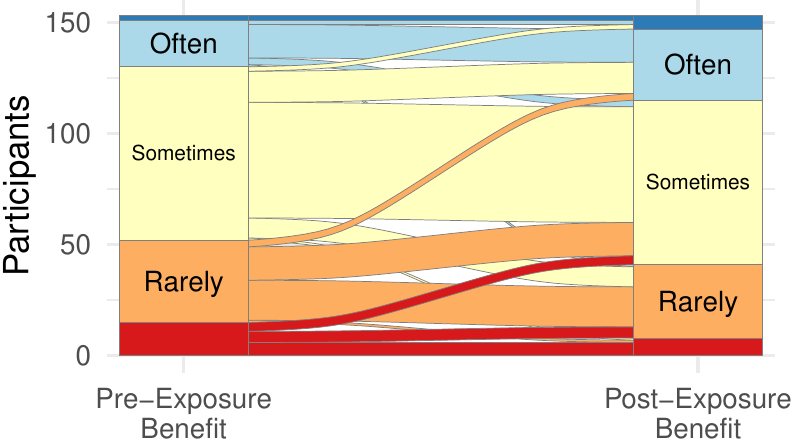}
    \caption[Frequency of benefit alluvium plot]{\label{fig:frequencyBenefit-alluvium}
        Frequency of benefit alluvium plot.
    }
\end{subfigure}
\caption[Alluvium plots of level of concern and frequency of benefit]{\label{fig:collectingConcern}
Detailed visualization of how the participants change their assessments of
\begin{enumerate*}[label=(\alph*)]
\item the level of concern (\ref{app:main-survey:Q2} \& \ref{app:main-survey:Q26}) and 
\item the frequency of benefit (\ref{app:main-survey:Q3} \& \ref{app:main-survey:Q27})
\end{enumerate*}
after interacting with the My Activity dashboard.
}
\end{figure*}
}

\newcommand{\figPaidPlanBar}[0]{
\begin{figure}[t]
\centering
\includegraphics[width=\columnwidth]{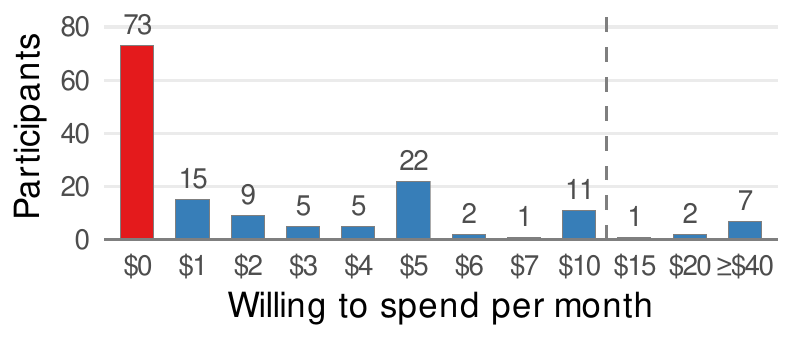}
\caption[Paid plan bar plot]{\label{fig:paidPlan-bar}
More than half of the participants ($n = \num{80}$; \SI{52.29}{\percent}) stated to pay at least \SI{1}[\dollar]{\USD} per month for Google services.
These are opposed by 73 (\SI{47.71}{\percent}) who would not pay any money.
Only \num{10} participants signalized a willingness to pay more than \SI{12}[\dollar]{\USD} per month (indicated by the dashed line).
}
\end{figure}
}

\newcommand{\figActivityAwarenessBar}[0]{
\begin{figure}[t]
\centering
\includegraphics[width=\columnwidth]{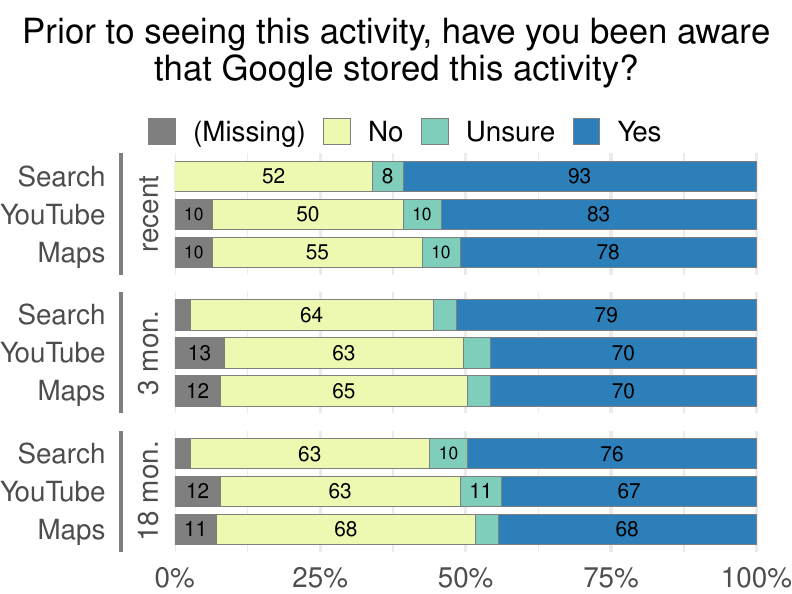}
\caption[Activity awareness bar plot]{\label{fig:activityAwareness-bar}
When presented with activities from their own My Activity feed, participants' awareness (\ref{app:main-survey:Q11}) seems to be similar regardless of the service.
The age of the activity however has small effect on the awareness (recent against 18 months).
}
\end{figure}
}

\newcommand{\figRememberActivityBar}[0]{
\begin{figure}[thbp]
\centering
\includegraphics[width=\columnwidth]{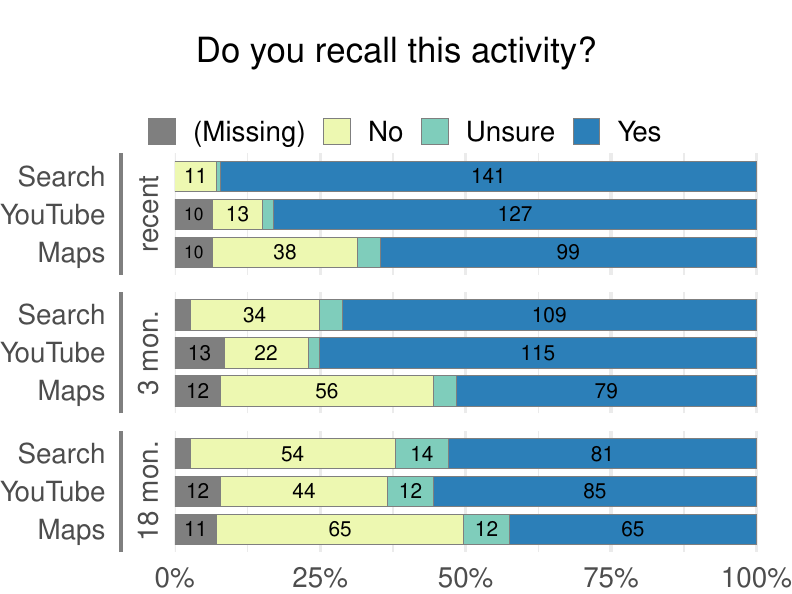}
\caption[Remember activity bar plot]{\label{fig:rememberActivity-bar}
The ability of participants to recall activities (\ref{app:main-survey:Q10}) decreases over time independent of activity type.
Google Maps activities in general seem to be harder to recall (Search / Maps: $W = \num{3480}$; $p < \num{0.001}$; $r = \num[round-mode=places]{0.25405121}$; YouTube / Maps: $W = \num{3609}$; $p < \num{0.001}$; $r = \num[round-mode=places]{0.30558110}$).
}
\end{figure}
}

\newcommand{\figCollectingActivityBar}[0]{
\begin{figure}[t]
\centering
\includegraphics[width=\columnwidth]{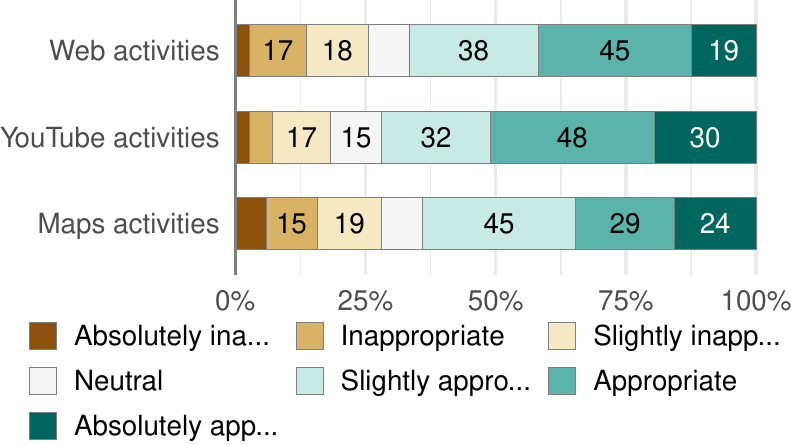}
\caption[Collecting activity is appropriate bar plot]{\label{fig:collectingActivity-bar}
The majority of participants found the explanations Google gives as to why they collect activity data appropriate (Web: \SI{66.67}{\percent}; YouTube: \SI{71.9}{\percent}; Maps: \SI{64.05}{\percent}).
}
\end{figure}
}

\newcommand{\figSettingsActivityBehaivorBar}[0]{
\begin{figure}[t]
\centering
\includegraphics[width=\columnwidth]{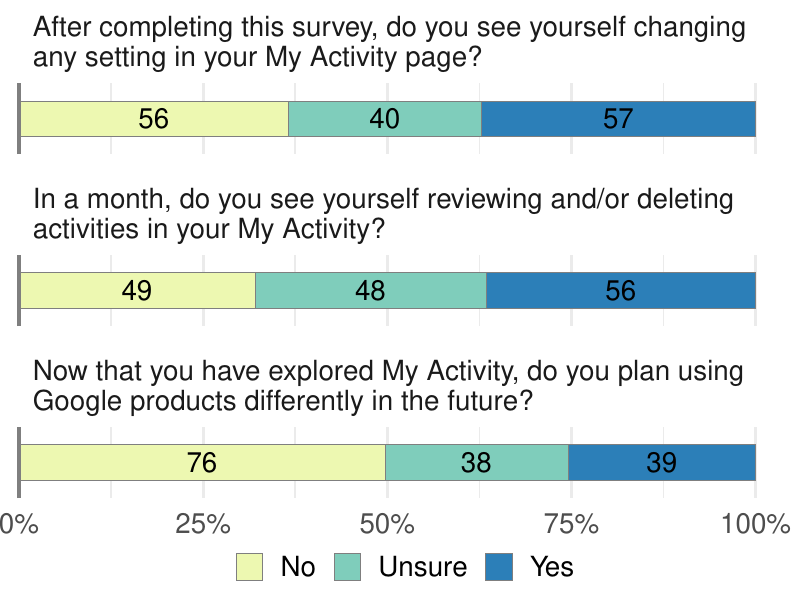}
\caption[Change Settings, Review Activities, and Change Behavior Bar Plot]{\label{fig:settingsActivityBehaivor-bar}
Willingness to take action after visiting the My Activity dashboard.
}
\end{figure}
}


\newcommand{\tabDemographics}[0]{
\begin{table}[t]
\centering
\small
\caption[Participant Demographics]{\label{tab:demographics}
Demographic data of the participants.
Age and gender data for our screening survey was provided by Prolific.
The IUIPC data was collected at the end of the screening survey.
Note: Prolific only provides binary gender data.
To get more precise data, we asked for gender and age at the end of the main study.
}
\renewcommand{\arraystretch}{0.6}
\begin{tabular*}{\columnwidth}{@{}>{\bfseries}ll@{\extracolsep{\fill}}*{4}{r}}
\toprule
&                & \multicolumn{2}{c}{\textbf{Screening}} & \multicolumn{2}{c}{\textbf{Main Study}} \\
&                & \multicolumn{2}{c}{\emph{(n = \num{669})}} & \multicolumn{2}{c}{\emph{(n = \num{153})}} \\
\midrule
\multirow{6}{*}{\rotatebox[origin=c]{90}{Gender}}
& & \multicolumn{1}{c}{\textbf{n}} & \multicolumn{1}{c}{\textbf{\%}} & \multicolumn{1}{c}{\textbf{n}} & \multicolumn{1}{c}{\textbf{\%}} \\
\cmidrule{3-6}
& Woman          & 317    & \pct{47.38} &  71    & \pct{46.41} \\
& Man            & 344    & \pct{51.42} &  79    & \pct{51.63} \\
& Non-binary     &  --    &          -- &   2    &  \pct{1.31} \\
& No answer      &   8    &  \pct{1.20} &   1    &  \pct{0.65} \\
\midrule
\multirow{6.5}{*}{\rotatebox[origin=c]{90}{Age}}
& 18--24         & 126    & \pct{18.83} &  29    & \pct{18.95} \\
& 25--34         & 152    & \pct{22.72} &  35    & \pct{22.88} \\
& 35--44         & 144    & \pct{21.52} &  31    & \pct{20.26} \\
& 45--54         & 128    & \pct{19.13} &  29    & \pct{18.95} \\
& 55+            & 116    & \pct{17.34} &  28    & \pct{18.30} \\
& No answer      &   3    &  \pct{0.45} &   1    &  \pct{0.65} \\
\midrule\multirow{7}{*}{\rotatebox[origin=c]{90}{IUIPC}}
& & \multicolumn{1}{c}{\textbf{Avg.}} & \multicolumn{1}{c}{\textbf{SD}} & \multicolumn{1}{c}{\textbf{Avg.}} & \multicolumn{1}{c}{\textbf{SD}} \\
\cmidrule{3-6}
& Control        &   5.8 &  1.0 &   5.9 &  1.0 \\
& Awareness      &   6.3 &  0.8 &   6.4 &  0.8 \\
& Collection     &   5.3 &  1.2 &   5.6 &  1.1 \\
\cmidrule{2-6}
& IUIPC Combined &   5.9 &  0.8 &   5.7 &  0.9 \\
\bottomrule
\end{tabular*}
\end{table}

}

\newcommand{\tabPostConcernRegression}[0]{
\begin{table}[t]
\centering
\caption[Post-exposure concern regression analysis]{\label{tab:post-concern-regression}
Ordinal regression model to describe the level of concern after visiting the My Activity dashboard.
The model uses a descending concern scale (\ie, from \emph{extremely} to \emph{not at all concerned}).
The Aldrich-Nelson pseudo $R^2$ of the model is \num[round-mode=places]{0.631858309728864}.
}
{
\small
\begin{tabular*}{\columnwidth}{
@{}
E{8.08926pt} 
@{\extracolsep{\fill}}
S[table-format = -2.2]
S[table-format = 3.2, scientific-notation = fixed, round-mode = places, round-minimum = 0.01, round-integer-to-decimal, tight-spacing = true] 
@{\extracolsep{0pt}}H@{\extracolsep{\fill}}
@{\extracolsep{0pt}}H@{\extracolsep{\fill}}
S[table-format = <1.3, scientific-notation = fixed, round-mode = places, round-precision = 3, round-minimum = 0.001]
@{\extracolsep{1pt}}
l
}
    \toprule
    {\textbf{Factor}} & {\textbf{Est.}} & {\textbf{OR}} & {\textbf{Error}} & {\textbf{t value}} & {\textbf{Pr(\textgreater\textbar z\textbar)}} & \\ 
    \midrule
    $\text{Pre-Exp. concern} = \textit{Extremely}$ & 5.71 & 302.8410699 & 1.03 & 5.56 & 2.74E-08 & *** \\ 
    $\text{Pre-Exp. concern} = \textit{Moderately}$ & 4.56 &  95.6011912 & 0.94 & 4.87 & 1.09E-06 & *** \\ 
    $\text{Pre-Exp. concern} = \textit{Somewhat}$ & 2.77 &  15.9609651 & 0.88 & 3.14 & 1.69E-03 & ** \\ 
    $\text{Pre-Exp. concern} = \textit{Slightly}$ & 1.18 &   3.2585878 & 0.86 & 1.38 & 1.67E-01 &  \\ 
    Increasing benefit & -0.17 &   0.8475816 & 0.37 & -0.45 & 6.54E-01 &  \\ 
    $\text{Knows My Activity} = Yes$ & -0.32 &   0.7247347 & 0.34 & -0.94 & 3.48E-01 &  \\ 
    $\text{IUIPC cont.} > 3.5$ & 0.29 &   1.3422555 & 1.07 & 0.27 & 7.84E-01 &  \\ 
    $\text{IUIPC awar.} > 3.5$ & -0.41 &   0.6607805 & 2.07 & -0.20 & 8.42E-01 &  \\ 
    $\text{IUIPC coll.} > 3.5$ & 0.29 &   1.3317022 & 0.54 & 0.53 & 5.95E-01 &  \\ 
    $\text{Gender} = Male$ & -0.24 &   0.7865547 & 0.34 & -0.71 & 4.81E-01 &  \\ 
    $\text{Age} \in\{18-34,~25-34\}$ & 0.39 &   1.4707258 & 0.48 & 0.81 & 4.17E-01 &  \\ 
    $\text{Age} \in\{35-44,~45-54\}$ & 0.91 &   2.4956446 & 0.47 & 1.95 & 5.15E-02 & $\cdot$ \\ 
    $\text{Edu.} \in \{\textit{No sch.g, (Sm.) HS}\}$ & 0.19 &   1.2054607 & 0.55 & 0.34 & 7.34E-01 &  \\ 
    $\text{Edu.} \in \{\textit{Sm. col., Assoc., Prof.}\}$ & 0.07 &   1.0735471 & 0.36 & 0.20 & 8.44E-01 &  \\ 
    $\text{Has IT background}$ & 0.64 &   1.9011462 & 0.40 & 1.62 & 1.05E-01 &  \\ 
    $\text{\# of activities} > \text{median}$ & -0.47 &   0.6269355 & 0.32 & -1.45 & 1.46E-01 &  \\ 
    \midrule
    \textbf{Intercepts} & & & & & \\
    \midrule
    $\textit{Not at all}~|~\textit{Slightly}$ & 0.58 &   1.7861953 & 2.01 & 0.29 & 7.73E-01 &  \\ 
    $\textit{Slightly}~|~\textit{Somewhat}$ & 2.87 &  17.6636336 & 2.03 & 1.41 & 1.58E-01 &  \\ 
    $\textit{Somewhat}~|~\textit{Moderately}$ & 4.63 & 102.0066655 & 2.05 & 2.26 & 2.40E-02 & * \\ 
    $\textit{Moderately}~|~\textit{Extremely}$ & 6.77 & 875.0841955 & 2.07 & 3.27 & 1.09E-03 & ** \\ 
    \bottomrule
\end{tabular*}
}
\footnotesize
\textbf{Signif. codes:} $\text{***}~\widehat{=} < 0.001$; $\text{**}~\widehat{=} <0.01$; $\text{*}~\widehat{=} <0.05$; $\cdot~\widehat{=} <0.1$
\end{table}

}

\newcommand{\tabPostBenefitRegression}[0]{
\begin{table}[t]
\centering
\caption[Post-exposure frequency of benefit regression analysis]{\label{tab:post-benefit-regression}
Ordinal regression model to describe the frequency of benefit after after visiting the My Activity dashboard.
In the model a ascending frequency scale (\ie, from \emph{never} to \emph{always}) is used.
The Aldrich-Nelson pseudo $R^2$ of the model is \num[round-mode=places]{0.681841649836333}.
}
{
\small
\newsavebox{\equbox}
\begin{tabular*}{\columnwidth}{
@{}
E{8.08926pt} 
@{\extracolsep{\fill}}
S[table-format = -2.2]
S[table-format = 1.2e+1, tight-spacing = true] 
@{\extracolsep{0pt}}H@{\extracolsep{\fill}}
@{\extracolsep{0pt}}H@{\extracolsep{\fill}}
S[table-format = <1.3, scientific-notation = fixed, round-mode = places, round-precision = 3, round-minimum = 0.001]
@{\extracolsep{1pt}}
l
}
    \toprule
    {\textbf{Factor}} & {\textbf{Est.}} & {\textbf{OR}} & {\textbf{Error}} & {\textbf{t value}} & {\textbf{Pr(\textgreater\textbar z\textbar)}} & \\ 
    \midrule
    $\text{Pre-Exp. benefit} = \textit{Never}$ & 22.29 & 4.81E+09 & 0.70 & 31.72 & 7.42E-221 & *** \\ 
    $\text{Pre-Exp. benefit} = \textit{Rarely}$ & 20.26 & 6.28E+08 & 0.51 & 39.44 & 2.22E-308 & *** \\ 
    $\text{Pre-Exp. benefit} = \textit{Sometimes}$ & 18.58 & 1.17E+08 & 0.52 & 35.95 & 5.37E-283 & *** \\ 
    $\text{Pre-Exp. benefit} = \textit{Often}$ & 16.12 & 1.00E+07 & 0.66 & 24.45 & 4.49E-132 & *** \\ 
    Increasing concern & 0.57 & 1.77E+00 & 0.51 & 1.11 & 2.68E-01 &  \\ 
    $\text{Knows My Activity} = Yes$ & -0.56 & 5.71E-01 & 0.37 & -1.50 & 1.33E-01 & \\  
    $\text{IUIPC cont.} > 3.5$ & 0.00 & 1.00E+00 & 1.02 & 0.00 & 9.98E-01 &  \\ 
    $\text{IUIPC awar.} > 3.5$ & 0.18 & 1.20E+00 & 2.24 & 0.08 & 9.35E-01 &  \\ 
    $\text{IUIPC coll.} > 3.5$ & 0.41 & 1.50E+00 & 0.57 & 0.72 & 4.73E-01 &  \\ 
    $\text{Gender} = Male$ & 0.92 & 2.51E+00 & 0.37 & 2.46 & 1.38E-02 & * \\ 
    $\text{Age} \in\{18-34,~25-34\}$ & 0.76 & 2.14E+00 & 0.50 & 1.53 & 1.26E-01 &  \\ 
    $\text{Age} \in\{35-44,~45-54\}$ & 0.58 & 1.79E+00 & 0.51 & 1.13 & 2.56E-01 &  \\ 
    $\text{Edu.} \in \{\textit{No sch.g, (Sm.) HS}\}$ & -0.16 & 8.50E-01 & 0.59 & -0.28 & 7.82E-01 &  \\ 
    $\text{Edu.} \in \{\textit{Sm. col., Assoc., Prof.}\}$ & -0.16 & 8.51E-01 & 0.38 & -0.42 & 6.71E-01 &  \\ 
    $\text{Has IT background}$ & -0.07 & 9.30E-01 & 0.42 & -0.17 & 8.64E-01 &  \\ 
    $\text{\# of activities} > \text{median}$ & 0.30 & 1.36E+00 & 0.34 & 0.89 & 3.73E-01 &  \\ 
    \midrule
    \textbf{Intercepts} & & & & & \\
    \midrule
    $\textit{Always}~|~\textit{Often}$ & 15.61 & 6.00E+06 & 1.68 & 9.28 & 1.69E-20 & *** \\ 
    $\textit{Often}~|~\textit{Sometimes}$ & 18.78 & 1.43E+08 & 1.63 & 11.52 & 1.05E-30 & *** \\ 
    $\textit{Sometimes}~|~\textit{Rarely}$ & 21.99 & 3.55E+09 & 1.66 & 13.27 & 3.32E-40 & *** \\ 
    $\textit{Rarely}~|~\textit{Never}$ & 24.56 & 4.63E+10 & 1.71 & 14.34 & 1.15E-46 & *** \\ 
    \bottomrule
\end{tabular*}
}
\footnotesize
\textbf{Signif. codes:} $\text{***}~\widehat{=} < 0.001$; $\text{**}~\widehat{=} <0.01$; $\text{*}~\widehat{=} <0.05$; $\cdot~\widehat{=} <0.1$

\end{table}

}

\newcommand{\tabGoogleProductsDifferentlyRegression}[0]{
\begin{table}[t]
\centering
\caption[Use Google differently regression analysis]{\label{tab:google-different-regression}
Binomial logistic model to describe which other factors 
influenced the participants plan to use Google products differently in the future (\emph{yes} responses to Question \ref{app:main-survey:Q25}).
The Aldrich-Nelson pseudo $R^2=\num[round-mode=places]{0.420131390603509}$.
}
{
\small
\begin{tabular*}{\columnwidth}{
@{}
E{8.08926pt} 
@{\extracolsep{\fill}}
S[table-format = -2.2]
S[table-format = 2.2e+1, tight-spacing = true]
@{\extracolsep{0pt}}H@{\extracolsep{\fill}}
@{\extracolsep{0pt}}H@{\extracolsep{\fill}}
S[table-format = <1.3, scientific-notation = fixed, round-mode = places, round-precision = 3, round-minimum = 0.001]
@{\extracolsep{1pt}}
l
}   
    \toprule
    {\textbf{Factor}} & {\textbf{Est.}} & {\textbf{OR}} & {\textbf{Error}} & {\textbf{z value}} & {\textbf{Pr(\textgreater\textbar z\textbar)}} & \\ 
    \midrule
    (Intercept) & -16.23 & 8.93E-08 & 3956.18 & -0.00 & 9.97E-01 &  \\ 
    Increasing concern & 2.50 & 1.22E+01 & 0.62 & 4.01 & 6.04E-05 & *** \\ 
    Increasing benefit & -1.31 & 2.71E-01 & 0.59 & -2.21 & 2.73E-02 & * \\ 
    $\text{IUIPC cont.} > 3.5$ & 18.42 & 1.00E+08 & 1361.72 & 0.01 & 9.89E-01 &  \\ 
    $\text{IUIPC awar.} > 3.5$ & -4.82 & 8.06E-03 & 4183.97 & -0.00 & 9.99E-01 &  \\ 
    $\text{IUIPC coll.} > 3.5$ & 1.40 & 4.07E+00 & 0.88 & 1.60 & 1.09E-01 &  \\ 
    $\text{Gender} = Male$ & -0.73 & 4.80E-01 & 0.51 & -1.45 & 1.46E-01 &  \\ 
    $\text{Age} \in\{18-34,~25-34\}$ & 0.97 & 2.63E+00 & 0.75 & 1.29 & 1.98E-01 &  \\ 
    $\text{Age} \in\{35-44,~45-54\}$ & 1.21 & 3.36E+00 & 0.76 & 1.60 & 1.09E-01 &  \\ 
    $\text{Edu.} \in \{\textit{No sch.g, (sm) HS}\}$ & 0.45 & 1.57E+00 & 0.78 & 0.58 & 5.62E-01 &  \\ 
    $\text{Edu.} \in \{\textit{Sm col. Assoc., Prof.}\}$ & -0.16 & 8.56E-01 & 0.50 & -0.31 & 7.57E-01 &  \\ 
    $\text{Has IT background}$ & 0.86 & 2.37E+00 & 0.55 & 1.58 & 1.14E-01 &  \\ 
    $\text{\# of activities} > \text{median}$ & -1.25 & 2.86E-01 & 0.46 & -2.72 & 6.49E-03 & ** \\ 
    \bottomrule
\end{tabular*}
}
\footnotesize
\textbf{Signif. codes:} $\text{***}~\widehat{=} < 0.001$; $\text{**}~\widehat{=} <0.01$; $\text{*}~\widehat{=} <0.05$; $\cdot~\widehat{=} <0.1$
\end{table}

}

\newcommand{\tabReviewActivitiesRegression}[0]{
\begin{table}[tbp]
\centering
\caption[Review activities regression analysis]{\label{tab:review-activity-regression}
Binomial logistic model to describe which other factors (beside visiting My Activity) influenced the participants plan to review/delete activities (\emph{yes} responses to Question \ref{app:main-survey:Q24}).
The  Aldrich-Nelson pseudo  $R^2 = \num[round-mode=places]{0.25321311230276}$.
}
{
\small
\begin{tabular*}{\columnwidth}{
@{}
E{8.08926pt} 
@{\extracolsep{\fill}}
S[table-format = -2.2]
S[table-format = 1.2e+1, tight-spacing = true]
@{\extracolsep{0pt}}H@{\extracolsep{\fill}}
@{\extracolsep{0pt}}H@{\extracolsep{\fill}}
S[table-format = <1.3, scientific-notation = fixed, round-mode = places, round-precision = 3, round-minimum = 0.001]
@{\extracolsep{1pt}}
l
}
    \toprule
    {\textbf{Factor}} & {\textbf{Est.}} & {\textbf{OR}} & {\textbf{Error}} & {\textbf{z value}} & {\textbf{Pr(\textgreater\textbar z\textbar)}} & \\ 
    \midrule
    (Intercept) & 13.77 & 9.53E+05 & 882.74 & 0.02 & 9.88E-01 &  \\ 
    Increasing concern & 0.90 & 2.45E+00 & 0.49 & 1.82 & 6.80E-02 & $\cdot$ \\ 
    Increasing benefit & 0.05 & 1.05E+00 & 0.41 & 0.13 & 9.00E-01 &  \\ 
    $\text{IUIPC cont.} > 3.5$ & 0.63 & 1.88E+00 & 1.25 & 0.51 & 6.13E-01 &  \\ 
    $\text{IUIPC awar.} > 3.5$ & -16.69 & 5.63E-08 & 882.74 & -0.02 & 9.85E-01 &  \\ 
    $\text{IUIPC coll.} > 3.5$ & 1.43 & 4.19E+00 & 0.71 & 2.03 & 4.24E-02 & * \\ 
    $\text{Gender} = Male$ & 0.45 & 1.56E+00 & 0.39 & 1.14 & 2.55E-01 &  \\ 
    $\text{Age} \in\{18-34,~25-34\}$ & -0.07 & 9.32E-01 & 0.53 & -0.13 & 8.93E-01 &  \\ 
    $\text{Age} \in\{35-44,~45-54\}$ & 0.24 & 1.27E+00 & 0.53 & 0.45 & 6.52E-01 &  \\ 
    $\text{Edu.} \in \{\textit{No sch.g, (Sm.) HS}\}$ & 0.85 & 2.34E+00 & 0.63 & 1.34 & 1.79E-01 &  \\ 
    $\text{Edu.} \in \{\textit{Sm. col., Assoc., Prof.}\}$ & -0.04 & 9.64E-01 & 0.41 & -0.09 & 9.28E-01 &  \\ 
    $\text{Has IT background}$ & 0.34 & 1.40E+00 & 0.44 & 0.77 & 4.41E-01 &  \\ 
    $\text{\# of activities} > \text{median}$ & -0.35 & 7.02E-01 & 0.36 & -0.97 & 3.30E-01 &  \\ 
    \bottomrule
\end{tabular*}
}
\footnotesize
\textbf{Signif. codes:} $\text{***}~\widehat{=} < 0.001$; $\text{**}~\widehat{=} <0.01$; $\text{*}~\widehat{=} <0.05$; $\cdot~\widehat{=} <0.1$
\end{table}

}

\begin{abstract}
Privacy dashboards and transparency tools help users review and manage the data collected about them online. Since 2016, Google has offered such a tool, My Activity, which allows users to review and delete their activity data from Google services.
We conducted an online survey with $n = \num{153}$ participants to understand if Google's My Activity, as an example of a privacy transparency tool, increases or decreases end-users' concerns and benefits regarding data collection.
While most participants were aware of Google's data collection, the volume and detail was surprising, but after exposure to My Activity, participants were significantly more likely to be both less concerned about data collection and to view data collection more beneficially.
Only \SI{25}{\percent} indicated that they would change any settings in the My Activity service or change any behaviors.
This suggests that privacy transparency tools are quite beneficial for online services as they garner trust with their users and improve their perceptions without necessarily changing users' behaviors.
At the same time, though, it remains unclear if such transparency tools actually improve end user privacy by sufficiently assisting or motivating users to change or review data collection settings.
\end{abstract}

\section{Introduction}\label{sec:intro}
Privacy dashboards~\cite{earp-16-my-activity, fischer-huebner-16-priv-dashboard, herder-20-privacy-dashboard} allow users of online services to review and control data collection.
Google introduced an activity dashboard called \emph{My Activity}~\cite{google-16-my-activity} in 2016 that allows users to view their activity history (such as searches, videos, and location data), turn off activity collection, and (automatically) delete activities from their history. 

While there has been research suggesting privacy dashboards~\cite{zimmermann-14-privacy-dashboards, fischer-huebner-16-priv-dashboard, raschke-17-usable-pri-dashboard, herder-20-privacy-dashboard} increase users' understanding of data collection, particularly around online behavioral advertising~\cite{ur-12-creepy, rader-14-awareness-google, chanchary-15-percep-tracking, weinshel-19-places-you-been, wei-20-twitter-knows} and interest inferences~\cite{tschantz-18-accu-google-ad, dolin-18-dd-inferences, rader-20-inferences-vs-perceptions}, there is little research on the impact of privacy dashboards on the perceived risks or benefits of the data collection itself. 

In this paper, we conducted an online survey with $n = \num{153}$ participants to explore how users' concerns of and benefits from Google's data collection are influenced by My Activity, as an exemplar privacy dashboard. Participants were first surveyed about their concern regarding Google's data collection and how frequently they benefit from it, both on a Likert scale and in open-ended responses.
They were then directed to their Google My Activity dashboard to view their own, real, activities that Google collected about them, and then participants were again asked about their concern/or benefit.
Through these methods, we were able to ask and answer the following research questions:
\begin{enumerate}[topsep=1pt, leftmargin=*, itemsep=1pt, 
label=\textbf{RQ\arabic*}, ref=\textbf{RQ\arabic*}]
    \item\label{RQ:awarness-and-understanding} \emph{[Awareness and Understanding]} What are users' awareness and understanding of Google's data collection?

    Participants are generally aware of and understand why Google collects activities, citing targeted advertising, personalization, and product improvements. However, while aware of the purposes, many express surprise with the volume and detail of activities. 

    \item\label{RQ:trans-tool-effect}  \emph{[Impact on Benefit/Concern]} How does the My Activity dashboard affect users' concern about and perceived benefit of Google's data collection? 

    Concern about Google's data collection significantly decreased, and perceived benefit significantly increased post exposure to My Activity, despite participants' qualitatively describing similar concerns and benefits before and after exposure.
    Ordinal logistic regression indicated that those who showed higher initial concern were much more likely to reduce their concern,
    and across all initial benefit levels, participants were almost always likely to increase their perceived benefit.

    \item\label{RQ:change-behavior} \emph{[Behavioral Change]} What settings and behaviors would users change due to exposure to My Activity?

    Most participants describe that they would not (\SI{37}{\percent}) or are unsure if (\SI{26}{\percent}) they would change any My Activity settings, and only \SI{25.49}{\percent} indicated that they plan to use Google products differently.
    Logistic regression suggests that those that had an increase in concern and decrease in benefit were much more likely ($\num{11.3}\times$ and $\num{2.1}\times$, respectively) to use Google differently.
\end{enumerate}

These results suggest that privacy dashboards and transparency tools are a net positive for online services.
Google's My Activity both decreases concerns about and increases perceived benefit of data collection, but it is not clear that these dashboards help end-users, broadly, to increase their privacy.
Most participants indicated that they would not use the features of the dashboard nor change their behavior. 

This may be because many users are already privacy resigned, believing that data collection will occur regardless of their choices, or it may be that the burden of properly managing their privacy is too high despite the availability of the transparency tool.
As more and more transparency tools become available, this burden will only increase, and so research into mechanisms to consolidate and automate management of data collection may greatly benefit users.

\section{Background: Google My Activity}\label{sec:myactivity}
Google introduced \emph{My Activity}\footnote{Google's My Activity, available at: \url{https://myactivity.google.com}, as of \today.} in June 2016~\cite{olivarez-giles-16-my-activity}, and it  enables users to manage their Google Web \& App, Location, and YouTube history and other data collected from Chrome, Android, etc.
My Activity is designed as a transparency tool, privacy dashboard, and data collection control mechanism and is the successor of Google's \emph{Web History}.

The My Activity pages offers a number of user benefits to data collection.
For example, ``\emph{more personalized experiences across all Google services},'' and it offers users ``\emph{faster searches, better recommendations},'' ``\emph{personalized maps, recommendations based on places you've visited},'' and ``\emph{better recommendations, remember where you left off, and more}.''\footnote{My Activity activity controls, available at: \url{https://myactivity.google.com/activitycontrols}, as of \today.}

My Activity lists \emph{activities} such as, ``Searched for \emph{USENIX 2021},''  and activity details , such as type of activity, timestamp, and device.
Viewed as a single event, bundle of events, or filtered by date ranges and services, users can review or delete activities, as well as enabled/disabled data collection and ad personalization.
Users receive a modal when disabling activity collection warning that this action will also disable personalization and not delete previously collected data.
(See \emph{Explore My Activity} section in \appref{app:main-survey} for a visual.)

In May 2019, Google added a setting to enable automatic deletion of activities (after 3 or 18~months)~\cite{miraglia-19-google-3-or-18-months}, and in August 2019, Google introduced an option to disable collecting audio recordings~\cite{bohn-20-google-voice-opt-out}.
In June 2020, Google updated their policy to give the option for auto-deleting activities during account creation for \emph{newly created} accounts after 18~months for Web \& App and Location activities and after 36~months for YouTube activities.
However, \emph{existing} accounts will still need to proactively enable the feature~\cite{newman-20-autodelete-18-months}.

\section{Related Work}\label{sec:related-work}
\paragraph{Online Behavioral Advertising.}
Many services track online activities of their users to infer interests for targeted advertising\,\cite{weinshel-19-places-you-been}.
There is much user-facing research on Online Behavioral Advertising~(OBA), including targeting and personalization\,\cite{wei-20-twitter-knows, hanson-20-hyper-personalize}, fingerprinting and tracking\,\cite{bielova-17-web-tracking, vastel-18-fp-scanner, degeling-19-take-cookies, iqbal-21-fingerprinters}, opting-out\,\cite{leon-12-johnny-opt-out, habib-19-opt-out, habib-20-scavenger-hunt, johnson-20-opt-out-adchoices-cost}, privacy-enhancing technologies\,\cite{schaub-16-watching-me, murmann-17-survey-tets, wu-18-private-browsing, coopamootoo-20-usage-pets}, usable privacy notices\,\cite{kelley-09-nutrition-label, schaub-15-effective-notices, gluck-16-short-notices}, cookie banners and consent\,\cite{utz-19-uninformed-consent, matte-20-cookie-banners, nouwens-20-consent-popups}, and also awareness, behaviors, perceptions, and privacy expectations\,\cite{malandrino-13-awareness, leon-13-willingness, rao-16-unexpected, acquisti-17-nudges, dolin-18-dd-inferences, rader-20-inferences-vs-perceptions}.
Below, we highlight some of the more related work.

Ur \etal\cite{ur-12-creepy} conducted interviews to explore non-technical users' attitudes about OBA, finding that participants were surprised that browsing history can be used to tailor advertisements.
Rader\,\cite{rader-14-awareness-google} studied users' awareness of behavioral tracking on Facebook and Google, suggesting that increased awareness of consequences of data aggregation led to increased concern.
Chanchary and Chiasson\,\cite{chanchary-15-percep-tracking} explored users' understanding of OBA and tracking prevention tools, noting that participants expressed more willingness to share data given control mechanism over collected data.
We find similarly in this study that My Activity is such a tool: Participants expressed decreased concern with data collection and were unlikely to change collection settings.

Most recently, Wei~\etal\,\cite{wei-20-twitter-knows} studied the advertising ecosystem of Twitter, exploring ad targeting criteria.
Similar to our work, participants shared some of their Twitter data via a browser extension.
The authors suggested that transparency regulations should mandate that the \enquote{right of access} not only includes access to the raw data files, but also a clear description and tools to visualize the data in a meaningful way.
My Activity provides such a meaningful way to visualize and access this data, but unfortunately, it still may not sufficiently motivate users to manage data collection.

\paragraph{Transparency and Privacy Dashboards.}
Transparency tools and privacy dashboards, which allow users to explore and manage data collection and privacy from online services, have been extensively proposed and explored in the literature~\cite{janic-13-tets-overview,raschke-17-usable-pri-dashboard,murmann-17-survey-tets,zimmermann-14-privacy-dashboards,rao-14-what-do-they-know,schnorf-14-transparency-trust,tschantz-18-accu-google-ad,weinshel-19-places-you-been,herder-20-privacy-dashboard,earp-16-my-activity}.
With the European General Data Protection Regulations (GDPR) (and other similar laws), data access requirements will likely lead to an increase in transparency tools and dashboards.
Below we outline some of the more related work.

Rao~\etal\,\cite{rao-14-what-do-they-know} suggested that dashboards were insufficient in providing transparency in to the creation of user profiles in a study of ad profiles from BlueKai, Google, and Yahoo, and as a result participants did not intend to change behaviors.
This same lack of transparency in My Activity may explain why many participants do not intend to change behaviors or settings.
Schnorf~\etal\,\cite{schnorf-14-transparency-trust} found that offering more control does not lead to less trust when exploring inferred interest  transparency tools, and we find similarly with My Activity.

Angulo \etal\,\cite{angulo-15-data-track} and Fischer-H\"ubner \etal\cite{fischer-huebner-16-priv-dashboard} developed \emph{Data Track}, a transparency tool for disclosing users data for different online services.
Tschantz \etal\,\cite{tschantz-18-accu-google-ad} compared inferred values displayed in Google's Ad Settings~\cite{google-09-ad-settings} to self-reported values, finding that logged in users were significantly more accurate.
Weinshel \etal\,\cite{weinshel-19-places-you-been} developed a browser extension that visualizes information that trackers could infer from browsing habits, greatly surprising users about the extent and prevalence of data collection.
We also see that while participants are aware of data collection, the scope of it is surprising to them.

Recently, Rader~\etal\,\cite{rader-20-inferences-vs-perceptions} investigated users' reactions to Google's and Facebook's profile inferences, and while many participants understood inferences to be a description of past activities, they were challenged to understand them as predictive of future interests and actions.
Rader~\etal argued for better transparency mechanisms, by adding explanations of how inferences might get used, and restricting inferences to only include the ones that can be explained by users, and thus, are not based on aggregation or inaccurate assumptions.
Meanwhile, Herder and van Maaren\,\cite{herder-20-privacy-dashboard} also found that removing derived and inferred data has a positive effect on trust and perceived risk.
Note that My Activity shows raw data, not inferred data, and it may be the case that better connecting specific inferences to data collection could improve transparency and better inform user choices.

Most related to our work, Earp and Staddon\,\cite{earp-16-my-activity} conducted a pilot study with about 100 undergraduate students on Google Ad Settings and Google Web History that---somewhat unfortunately---was rebuilt and became Google My Activity during their data collection in 2016.
For the participants that had ``sufficient'' data accessible, they found no evidence that the tools were harmful to user trust and privacy.
Our work confirms this finding, and goes further by showing that My Activity can be helpful in reducing concerns and increasing perceived benefits for end users.
Additionally, as My Activity has been active for 4--5 years at the time of our study, our work is able to explore the impact of this transparency tool.

\section{Method}\label{sec:study}

We designed our study for participants to directly interact with their own activity history on My Activity, following a pre-post-study design.
First, participants answered questions regarding their concern for and benefit from Google's data collection, and after exposure to My Activty, they answered the same set of questions.
In the rest of this section, we outline our study protocol, recruitment, limitations, and ethical considerations.

\figstudydesign{}

\subsection{Study Procedure}
To ensure that participants had active Google accounts, we used a two-part structure with a \emph{screening survey} where qualified participants were asked to participate in the \emph{main study}.
The full screening survey can be found in \appref{app:pre-survey}, and the main study can be found in \appref{app:main-survey}.

\paragraph{Screening Survey.}
We used the following inclusion criteria to screen participants for the main study:
\begin{enumerate*}[label=(\roman*)]
\item the participant has an active Google account, 
\item the participant has used their Google account for more than three years, 
\item the participant currently uses Google Search, Google Maps, and YouTube.
\end{enumerate*}

In the screening survey we also asked participants if they have a Gmail account (as surrogate for a Google account), the age of the account, and what other Google products (besides Gmail) they use and their frequency of use and overall importance.
Participants also answered the Internet users' information privacy concerns (IUIPC) questionnaire, as described by Malhotra~\etal\cite{malhotra-2004-iuipc}, to gain insights into participants' privacy concerns.

\paragraph{Main Study.}
If participants qualified they were invited to complete the main study which is divided into three stages:
\begin{enumerate*}[label=(\roman*)]
\item a pre-exposure stage, in which participants install the survey browser extension that aided in administering the survey and answer questions about their perceptions of Google;
\item an intervention stage consisting of two steps;
\begin{enumerate*}[label=(\alph*)]
\item an exploration phase step and
\item an activity presentation step
\end{enumerate*}
\item a post-exposure stage.
\end{enumerate*}
To facilitate the study, we designed a custom browser extension that locally analyzes My Activity to collect aggregated information about the number of activities of users and also to fill-in survey questions.
Participants are given detailed instructions to both install and uninstall the extension.

Below, we describe each part of the study in detail (see \autoref{fig:study-design} for a visual).
\begin{enumerate}[leftmargin=*,noitemsep]
    
    \item Informed Consent: Participant consented to the study; the consent included that participants would be asked to install a web browser extension and answer questions about their experience with Google's My Activity page.
    
    \item Install Extension: Participants installed the browser extension that assisted in administering the survey.
    The extension also recorded aggregate information about the survey participants' number of activities per month for each activity category (\eg, Google Search, YouTube) and the date of the oldest activity, as a proxy for account age.
    
    \item Pre-Exposure Perceptions of Google: Participants were asked about their awareness of Google's data collection practices, their level of concern, and how often they benefit from Google's collection of their online activities, both on a Likert scale and in open-ended responses.
    We also asked participants if they employed any strategies to limit the amount of data that Google may collect about them.
    The questions about perceived level of concern and benefit serve as a pre-exposure baseline and are asked again after exposure to the Google My Activity page and recent/historical Google activities.
    Questions:~\ref{app:main-survey:Q1}--\ref{app:main-survey:Q4}.
    
    \item Visit My Activity: We provided participants with a brief descriptive introduction to the My Activity service and the term ``activities'' as used by Google.
    Participants were presented with a ``Sign in with Google'' button and were instructed to login to their primary Google account.
    Then participants explored their My Activity for two minutes, managed by the browser extension with an overlay banner and restricting navigation away from My Activity.
    After two minutes, participants were directed back to the survey.

    \item My Activity Questions: Participants were asked to provide their immediate reactions to My Activity and their reasoning for why Google is collecting this data.
    Participants were also asked if they perceive the data collection to be beneficial or harmful, if they have any concerns, and whether this data collection improves their experience using Google services.
    Questions:~\ref{app:main-survey:Q5}--\ref{app:main-survey:Q9}.
    
    \item Activity Presentation: Next the browser extension locally displayed three recent activities (randomly selected from \numrange{2}{12}~days old), three three-month-old activities (randomly selected from \numrange{90}{100}~days old), and three 18-month-old activities (randomly selected from \numrange{540}{550}~days old).
    The participants reported their awareness and recall of each of the nine activities, which were selected with an even distribution from the services Google Search, YouTube, and Google Maps.
    Questions:~\ref{app:main-survey:Q10}--\ref{app:main-survey:Q14}.
    
    \item Reflection and Trust:
    We then asked the participants to reflect on their post-exposure feelings and on the appropriateness of the data collection.
    Questions:~\ref{app:main-survey:Q15}--\ref{app:main-survey:Q19}.
    
    \item Change Behavior: Participants were asked what behavioral change they would likely implement after learning about My Activity,
    if they planned to change how long Google stores their activities, or if they would like to delete their activities.
    Participants were also asked if they plan to change their My Activity settings and if they would interact differently with Google products in the future.
    Questions:~\ref{app:main-survey:Q20}--\ref{app:main-survey:Q25}.
    
    \item Post-Exposure Perception of Google: We again asked participants about their concern for and benefit from Google's data collection.
    Questions \ref{app:main-survey:Q26}, \ref{app:main-survey:Q27}.
    
    \item Demographics: Participants were asked to provide demographic information, such as age, identified gender, education, and technical background.
    Questions:~\ref{app:main-survey:D1}--\ref{app:main-survey:D4}.
    
    \item Uninstall Extension: Upon completing the survey participants were
    instructed to remove the browser extension.
\end{enumerate}

\subsection{Recruitment and Demographics}
We recruited \num{669}~participants via \emph{Prolific}\footnote{\url{https://www.prolific.co} - Prolific participant recruitment service, as of \today.} for the screening survey.
After applying our inclusion criteria, \num{447} participants qualified for the main study.
Of those, \num{153} completed the main study; unfortunately, rates of return to the main study fell below \SI{50}{\percent}.
On average, it took \num{4} minutes for the screening survey and \num{26} minutes for the main study. Participants who completed the screening survey received \SI{0.50}[\dollar]{\USD} and \SI{3.75}[\dollar]{\USD} for completing the main study.

We sought a balanced recruitment between gender and five age ranges (\numrange[range-phrase = --]{18}{24}, \numrange[range-phrase = --]{25}{34}, \numrange[range-phrase = --]{35}{44}, \numrange[range-phrase = --]{45}{54}, \num{55}+) with a median participant age of \num{38}.
Purposive sampling was performed using Prolific's built in study inclusion criteria which allows researchers to specify availability based on Prolific's pre-screened demographics.
The identified gender distribution for the main study was \SI{52}{\percent} men, \SI{46}{\percent} women, and \SI{2}{\percent} non-binary or did not disclose gender.
Participant demographics are presented in \autoref{tab:demographics} (additional demographic information can be found in \appref{app:demographics}).

\begin{table}[t]
\centering
\small
\caption[Participant Demographics]{\label{tab:demographics}
Demographic data of the participants.
Age and gender data for our screening survey was provided by Prolific.
The IUIPC data was collected at the end of the screening survey.
Note: Prolific only provides binary gender data.
To get more precise data, we asked for gender and age at the end of the main study.
}
\renewcommand{\arraystretch}{0.6}
\begin{tabular*}{\columnwidth}{@{}>{\bfseries}ll@{\extracolsep{\fill}}*{4}{r}}
\toprule
&                & \multicolumn{2}{c}{\textbf{Screening}} & \multicolumn{2}{c}{\textbf{Main Study}} \\
&                & \multicolumn{2}{c}{\emph{(n = \num{669})}} & \multicolumn{2}{c}{\emph{(n = \num{153})}} \\
\midrule
\multirow{6}{*}{\rotatebox[origin=c]{90}{Gender}}
& & \multicolumn{1}{c}{\textbf{n}} & \multicolumn{1}{c}{\textbf{\%}} & \multicolumn{1}{c}{\textbf{n}} & \multicolumn{1}{c}{\textbf{\%}} \\
\cmidrule{3-6}
& Woman          & 317    & \pct{47.38} &  71    & \pct{46.41} \\
& Man            & 344    & \pct{51.42} &  79    & \pct{51.63} \\
& Non-binary     &  --    &          -- &   2    &  \pct{1.31} \\
& No answer      &   8    &  \pct{1.20} &   1    &  \pct{0.65} \\
\midrule
\multirow{6.5}{*}{\rotatebox[origin=c]{90}{Age}}
& 18--24         & 126    & \pct{18.83} &  29    & \pct{18.95} \\
& 25--34         & 152    & \pct{22.72} &  35    & \pct{22.88} \\
& 35--44         & 144    & \pct{21.52} &  31    & \pct{20.26} \\
& 45--54         & 128    & \pct{19.13} &  29    & \pct{18.95} \\
& 55+            & 116    & \pct{17.34} &  28    & \pct{18.30} \\
& No answer      &   3    &  \pct{0.45} &   1    &  \pct{0.65} \\
\midrule\multirow{7}{*}{\rotatebox[origin=c]{90}{IUIPC}}
& & \multicolumn{1}{c}{\textbf{Avg.}} & \multicolumn{1}{c}{\textbf{SD}} & \multicolumn{1}{c}{\textbf{Avg.}} & \multicolumn{1}{c}{\textbf{SD}} \\
\cmidrule{3-6}
& Control        &   5.8 &  1.0 &   5.9 &  1.0 \\
& Awareness      &   6.3 &  0.8 &   6.4 &  0.8 \\
& Collection     &   5.3 &  1.2 &   5.6 &  1.1 \\
\cmidrule{2-6}
& IUIPC Combined &   5.9 &  0.8 &   5.7 &  0.9 \\
\bottomrule
\end{tabular*}
\end{table}

{}

\subsection{Analysis Methods and Metrics}

\paragraph{Qualitative Coding.}
We conducted qualitative open coding to analyze \num{19} free-response questions.
A primary coder from the research team crafted a codebook and identified descriptive themes by coding each question.
A secondary coder coded a \SI{20}{\percent} sub-sample from each of the free-response questions over several rounds, providing feedback on the codebook and iterating with the primary coder until inter-coder agreement was reached (Cohen's $\kappa > 0.7$).
We report the number of responses receiving a code and percentage of responses assigned that code.
Note that responses may be assigned multiple codes.

\paragraph{Statistical Tests and Regression Analysis.}
We performed two Wilcoxon signed-rank tests for repeated measurements on the Likert responses to the pre and post-exposure questions on concern (\ref{app:main-survey:Q2}, \ref{app:main-survey:Q26}) and benefit (\ref{app:main-survey:Q3}, \ref{app:main-survey:Q27}).
The same tests were used to find differences between the responses \ref{app:main-survey:Q11}--\ref{app:main-survey:Q14} for the presented activities, and then post-hoc, pairwise analysis using again Wilcoxon signed-rank tests between categories, with Holm correction for overlapping measures.

We also performed two proportional odds logistic regressions to analyze which factors, in addition to the intervention, that may have influenced the Likert responses moving up or down the scales for concern (\ref{app:main-survey:Q26}) and benefit  (\ref{app:main-survey:Q27}).

Finally, we performed three binomial logistic regressions on behavior change questions: Google settings~\ref{app:main-survey:Q23}, review/delete activities~\ref{app:main-survey:Q24}, and  use Google products differently in the future~\ref{app:main-survey:Q25}.
Since we were interested whether participants planned to take action, we binned the \emph{unsure} and \emph{no} responses.

\subsection{Ethical Considerations}
The study protocol was approved by our Institutional Review Board (IRB) with approval number NCR202596, and throughout the process, we considered the sensitivity of participants' My Activity data at every step.
At no point did (do) the researchers have access to participants' precise Google activities.
All aspects of the survey requiring access to actual Google activity was administered \emph{locally} on the participant's machine using the browser extension.
We did not collect information about individual activities to protect participants privacy, and only report information in aggregate, \eg, the number of activities per month.
All participants were informed about the nature of the study prior to participating and consented to participating in both the screening and main study.
At no time did the extension nor the researchers have access the participants' Google password or to any other Google account data, and all collected data is associated with random identifiers.

\subsection{Limitations}
Our study is limited in its recruitment, particularly to Prolific users residing in the U.S.
We attempted to compensate by performing purposive sampling on Prolific to balance demographic factors like age and gender, but we cannot claim full generalizability of the results.
Despite this limitation, prior work~\cite{redmiles-19-generalize} suggests that online studies about privacy and security behavior can approximate behaviors of populations.

Social desirability and response bias may lead to participants over describing their awareness of Google data collection as they may believe that this is the expectation of the researchers.
Such biases may be most present when participants indicate if they intend to change a setting or behavior.

Our regression analysis is, unfortunately, under-powered to identify small effects as we only have 153 examples.
However, the pseudo $R^2 > 0.5$ for the ordinal-logistic regression, suggesting excellent fit; the logistic regressions have pseudo $ 0.25 < R^2 < 0.68$, also suggesting good fits.
As a result, we have confidence that the models are describing meaningful covariants, but small effects may not be captured.

Finally, as a pre-post-study we attribute changes in concern and benefit to the intervention, namely exposure to My Activity, but we cannot rule out other factors impacting changes in concern and benefit.
A randomized control trial would be needed to completely rule out other factors, but using such a methodology here is unclear because there is limited control of the display of activities and behaviors of our online participants outside of the study.

\section{Results}\label{sec:results}
This section is structured along our research questions.
We first present our findings concerning the participants' awareness and understanding of Google's data collection practices.
Secondly, we show the impact of Google's My Activity on the perceived concern and benefit of the participants.
Finally, we discuss what actions participants plan to take as a result of interacting with My Activity.

\figActivityAwarenessBar{}

\subsection{\ref{RQ:awarness-and-understanding}: Awareness and Understanding}
As part of \ref{RQ:awarness-and-understanding}, we seek to understand if participants are aware of Google's My Activity, understand the scope of Google's data collection and how that data is used.

\figRememberActivityBar{}

\paragraph{Prior Awareness of My Activity.}
Even though Google introduced My Activity in 2016, only a third ($n = \num{55}$; \SI{35.9}{\percent}) of the participants indicate that they have visited their My Activity page prior to our study.
We also asked the participants to assess how aware they were of Google's practice to collect data on individuals' use of their services.
This first question served---together with the Questions \ref{app:main-survey:Q2} and \ref{app:main-survey:Q3} (see \appref{app:main-survey})---to get a first impression of participants' attitudes towards data collection and privacy.
Most participants ($n = \num{115}$; \SI{75.2}{\percent}) indicated they were at least \emph{somewhat aware} ($n = \num{42}$; \SI{27.5}{\percent}), \emph{moderately aware} ($n = \num{54}$; \SI{35.3}{\percent}), or even \emph{extremely aware} ($n = \num{19}$; \SI{12.4}{\percent}).
Only \num{6} (\SI{3.92}{\percent}) participants stated they were \emph{not at all aware}.

\paragraph{Privacy Management Strategies.} Qualitative coding of \ref{app:main-survey:Q4} indicates a divide between the participants who attempt to apply a specific privacy management strategy and those who appear to be privacy resigned or unconcerned, and thus do not have a management strategy.
For instance:
\begin{quote}
    \small
    \emph{
        No strategies.
        I just use Chrome and whatever information Google gets they get.
        I signed up and accepted that they would take my data and information.
    } (P61)
\end{quote}
\begin{quote}
    \small
    \emph{
        No, I don't.
        I don't mind that they collect data about my usage and statistics.
    } (P21)
\end{quote}

Half of the participants ($n = \num{78}$; \SI{50.98}{\percent}) claimed not to have strategies for managing the kind of information Google may collect about them, while 38 (\SI{24.8}{\percent}) participants explained that they employed web browser based strategies such as opening private or incognito windows ($n = \num{17}$; \SI{11.1}{\percent}), installing ad-blocking or tracking prevention browser extensions ($n = \num{10}$; \SI{6.54}{\percent}), and clearing their browser history or cookies ($n = \num{9}$; \SI{5.88}{\percent}).
Others indicated that they limit the information that they provide ($n = \num{25}$; \SI{16.3}{\percent}), limit their usage of Google products or refrain from logging into their Google accounts ($n = \num{7}$; \SI{4.58}{\percent}), provide false information ($n = \num{6}$; \SI{3.92}{\percent}), or delete information ($n = \num{3}$; \SI{1.96}{\percent}).

\paragraph{Scope of Data Collection.}
We asked a set of free-response questions after the participants visited their My Activity page to gauge immediate reactions (\ref{app:main-survey:Q5}).
One-third ($n = \num{51}$; \SI{33.3}{\percent}) of study participants' immediate reaction was that of surprise, \eg, \enquote{I am surprised at how much of my browsing activity is saved and is identifiable} (P72), and \enquote{It's an awful lot of my life on that page} (P11).
Furthermore, 54 (\SI{35.3}{\percent}) participants stated that the amount of data collected on the My Activity page was more than they expected.
For example:
\begin{quote}
    \small
    \emph{
        I'm surprised at how much data google collects beside it's own sites.
        I did not know it saved the links you clicked on after a google search, for instance.
    } (P23)
\end{quote}

Others were not surprised ($n = \num{34}$; \SI{22.2}{\percent}) and stated the amount of data collection was as expected ($n = \num{30}$; \SI{19.6}{\percent}).
For instance:
\begin{quote}
    \small
    \emph{
        It didn't surprise me to see a tracking of all of my activity.
        Perhaps it gives me a way to control the information tracking in the future.
    } (P89)
\end{quote}

Some participants found the My Activity page helpful ($n = \num{16}$; \SI{10.5}{\percent}) and were interested ($n = \num{9}$; \SI{5.88}{\percent}), while a few participants reacted with concern ($n = \num{6}$; \SI{3.92}{\percent}), felt uncomfortable ($n = \num{4}$; \SI{3.92}{\percent}), or thought it creepy ($n = \num{4}$; \SI{2.61}{\percent}).

This is in line with closed responses to awareness of data collection types for individual activities (\ref{app:main-survey:Q11}); as \autoref{fig:activityAwareness-bar} shows, for recent search activities \SI{60.78}{\percent} of the participants indicated awareness.
For 18-month-old YouTube activities, only \SI{43.79}{\percent} of the participants responded with \emph{yes}.
Comparing across services and activity ages, we find that there  is a significant difference between awareness of recent and 18-month old activities ($W = \num{1511}$; $p = \num{0.004}$; $r = \num[round-mode=places]{0.16678491}$).

Note that not all participants had activities for each combination of services and time frames (see missing data in \autoref{fig:activityAwareness-bar} and \ref{fig:rememberActivity-bar}).
For \num{24} participants, we could not obtain a full set of nine activities, \num{14} participants saw six activities during the survey, and six participant had seven activities.
One participant saw only one activity and the remaining three participants saw two, three, or eight activities.

\autoref{fig:rememberActivity-bar} shows the results of \ref{app:main-survey:Q10}.
The participants report higher recall for recent activities compared to older ones (recent / 3 months: $W = \num{1711}$; $p < \num{0.001}$; $r = \num[round-mode=places]{0.2589939}$; recent / 18 months: $W = \num{1862}$; $p < \num{0.001}$; $r = \num[round-mode=places]{0.4835289}$; 3 months / 18 months: $W = \num{3062}$; $p < \num{0.001}$; $r = \num[round-mode=places]{0.2887735}$).
Around half of the participants were able to recall their 18-month-old Search ($n = \num{81}$; \SI{52.94}{\percent}) or YouTube activities ($n = \num{85}$; \SI{55.56}{\percent}).
For Maps activities the fraction was even lower ($n = \num{65}$; \SI{42.48}{\percent}).
In contrast, \SI{92.16}{\percent} ($n = \num{141}$) of the participants could remember their recent Google Search activities.
However, even recent Google Maps activities were harder to recall for the participants ($n = \num{99}$; \SI{64.71}{\percent} could recall them).
Compared with recent Google Search activities, there is a significant difference with a large effect size ($W = \num{2643.5}$; $p < \num{0.001}$; $r = \num[round-mode=places]{0.6544004}$).

We assume this difference is due to the fact that some of the Google Maps activities were generated without the participants actively interacting with the service while Search activities are basically queries made via Google Search.

Note that not all participants had activities for all services and time periods.
In total \num{76} (of \num{1377}) records for the activity presentation of \num{24} participants were missing.

\figbetterUnderstandingBar{}

\paragraph{Understanding of Data Collection.}
We also recorded the mouse movements of the participants during their visit of the My Activity page to get an idea of whether and how they interacted with the page.
We recorded an average participant scroll depth of \num{20553}~pixels ($SD=\num{22285}$, $min=\num{657}$, $max=\num{252735}$).
A single activity height is approximately \num{200}~pixels, which suggests that the average participant scrolled past approximately \num{100}~activities during their exploration.

Asked whether My Activity helps to better understand what data Google collects, most participants ($n = \num{116}$; \SI{76.32}{\percent}) agreed.
Only \SI{12.42}{\percent} ($n = \num{19}$) indicated that it did not help.
\autoref{fig:betterUnderstanding-bar} shows the full results of this question.
And when asked to explain why they think My Activity helps them to better understand what data Google collects (\ref{app:main-survey:Q23a}), 61 (\SI{39.9}{\percent}) participants reported that My Activity provides transparency about the collected data, \eg, \enquote{I didn't realize some of this info was collected} (P4), and 

\figCollectingActivityBar{}

\figConcernAndBenefitBar{}

\begin{quote}
    \small
    \emph{
        I see what they are collecting.
        I feel like I always knew they were watching every site I visited but to quantify it gives me a better understanding.
    } (P66)
\end{quote}

Still other participants ($n = \num{31}$; \SI{20.3}{\percent}) were skeptical and felt the My Activity page did not show all the data Google collects, \eg, \enquote{I see the data that they are retaining, but I'm concerned that there is more data being saved that they're not sharing with me} (P148), and
\begin{quote}
    \small
    \emph{
        I think it gives me a better understanding, but I don't believe Google is being completely transparent on their end with what they keep or use.
        It is just what I can control on my end.
    } (P69)
\end{quote}

For some participants ($n = \num{13}$; \SI{8.497}{\percent}) My Activity did not help them better understand what data Google collects.
For example:
\begin{quote}
    \small
    \emph{
        It shows me what I have done but not how they are using it or what they are collecting from this data.
        Like are they collecting what I do in the app, what I engage with, how long I'm there what keeps my interest.
    } (P17)
\end{quote}

\paragraph{Purpose of Data Collection.}
We asked the participants to think of three purposes for which Google might collect this data (\ref{app:main-survey:Q7}).
Most participants ($n = \num{123}$; \SI{80.4}{\percent}) stated that the purpose for the collection was targeted advertising.
For example: \enquote{Make advertisements more targeted and effective} (P22), and \enquote{To target advertisements at me from my search history} (P29).
The next largest group identified experience improvements that include personalization ($n = \num{109}$; \SI{71.2}{\percent}) as the purpose, \eg, \enquote{Customize my search results based on interest} (P39), and product improvements ($n = \num{42}$; \SI{24.5}{\percent}), \eg, \enquote{Usage data for company research for products and programs} (P149).
Some participants ($n = \num{59}$; \SI{38.6}{\percent}) thought that Google's purpose was to sell their usage data.
P10 said, \enquote{Sell my data to third parties for profit,} and P31 said, \enquote{To sell to other companies.}

The purposes provided are mostly in line with what Google describes on its help pages, where they indicate the following reasons to collect activity data:
\begin{enumerate*}[label=(\roman*)]
    \item product improvements,
    \item recommendations,
    \item personalizations, and
    \item browser/search/location history.
\end{enumerate*}
However, knowing the purpose for the data does not imply agreement with the use, and so we also presented participants with Google's explanations for data collection, asking participants to gauge appropriateness of the explanation (\ref{app:main-survey:Q17}-\ref{app:main-survey:Q19}).
For all three activity categories, \autoref{fig:collectingActivity-bar} shows that \SI{64.05}{\percent} think the reasons to collect activity data are at least \emph{slightly appropriate}.

\subsection{\ref{RQ:trans-tool-effect}: Impact on Benefit \& Concern}
Google's My Activity dashboard provides extensive insights into data collection, and in this research question we seek to understand if exposure to My Activity affects concerns about or beliefs in benefits of Google's data collection.
We evaluate two Likert questions, one about concern (\ref{app:main-survey:Q2}, \ref{app:main-survey:Q24}) and one about benefit (\ref{app:main-survey:Q3}, \ref{app:main-survey:Q25}), before and after exposure to My Activity, as well as open-response explanations in answering this research question.
The responses are visualized in Figure~\ref{fig:concern-and-benefit}.

\figConcernAndBenefitAlluvium{}

\paragraph{Initial Perceptions Concerns.}
When participants were asked to explain their concern (\ref{app:main-survey:Q2a}) with the amount of information Google is collecting, more than half of the participants ($n = \num{79}$; \SI{51.6}{\percent}) said they had privacy concerns, such as concerns about the amount of information ($n = \num{15}$), \eg, \enquote{I feel like Google is taking way too much of my data} (P128), sensitivity of the information ($n = \num{14}$), \eg, \enquote{I'm concerned that the data collected can be very specific and in turn, identifying} (P103), and feeling uncomfortable sharing information ($n = \num{12}$), \eg, \enquote{My information is private and should be shared with no one} (P54).
For some participants ($n = \num{29}$; \SI{18.95}{\percent}) the unknowns were concerning, such as how the information is used ($n = \num{19}$), and who has access to the information collected ($n = \num{5}$).
For example P95 said, \enquote{I don't know what is being done with my personal information that Google collects and who is capable of gaining access to it.}
Security was also a concern for some participants ($n = \num{22}$; \SI{14.4}{\percent}), specifically concerns about data misuse ($n = \num{18}$) and personal data being released ($n = \num{8}$).
This quote from P138 is an example: \enquote{I am concerned about any platform, application or website wrongfully accessing my data or having a breach of the data I provide.}
Still others ($n = \num{16}$; \SI{10.5}{\percent}) responded that there existed a trade-off between privacy and free services, such as P115 who said: \enquote{I don't like that my privacy is being compromised, but overall I enjoy the convenience of the services and feel its worth it.}

\paragraph{Initial Perceptions of Benefits.}
When explaining the benefit (\ref{app:main-survey:Q3a}), participants described the benefits of improved suggestions ($n = \num{46}$; \SI{30.1}{\percent}), personalized advertisements ($n = \num{24}$; \SI{15.7}{\percent}), and the availability of usage history ($n = \num{15}$; \SI{9.8}{\percent}).
For example, P11 said, \enquote{I'm given information and predictions about what I'm looking for in a more precise and efficient manner, because my data has clued Google in.}
Participant P39 who found personalized advertisements useful said, \enquote{I receive ads that I have interest in and do not see `annoying' ads as a result.} Participant P26 had this to say about the benefits data collection for usage history: \enquote{Use of My Activity helps me retrace my steps and find information that I may need at a later date.}
Other participants ($n = \num{26}$; \SI{17}{\percent}) said they perceived no benefit, such as participant P\num{17}, who said \enquote{All they do is bombard me with more ads and it doesn't help me to do anything.}

\paragraph{Other Concerns.}
We additionally asked if participants had other concerns (\ref{app:main-survey:Q8}) prior to exposure, and many participants reported privacy concerns ($n = \num{58}$; \SI{37.9}{\percent}), security concerns ($n = \num{31}$; \SI{20.3}{\percent}), and too many unknowns ($n = \num{22}$; \SI{14.4}{\percent}).
Among the privacy concerns were concerns about selling information ($n = \num{14}$) and third parties ($n = \num{12}$); for instance, participant 19 had this to say:

\begin{quote}
    \small
    \emph{
        Google sells my information as a product.
        I am not really a customer.
        I am like a piece of corn that is sold on the commodities market.
        The farmer, Google, feeds my information and I respond.
        I am then sold to the highest bidder several times.
    } (P19)
\end{quote}

There were also privacy concerns about the amount of information ($n = \num{12}$), \eg, \enquote{It's just an odd feeling, knowing they collect every bit of information about me and keep it probably forever.} (P108)

Participants' security concerns were about data breach ($n = \num{29}$), \eg, \enquote{It does worry me if they ever had a data breach because it seems like they do have a lot of information about their users} (P143), and potential data misuse ($n = \num{18}$) \eg, \enquote{I also worry about hacking and unsavory entities using my information in ways I don't even understand} (P89), and \enquote{I am confident that given the opportunity, some human with any access to the data will use it for selfish reasons, possibly to the detriment of others} (P127).

\paragraph{Changing Level of Concern.}
To determine if there are significant changes in perceived concerns, we performed a Wilcoxon signed-rank test on the Likert responses before (\ref{app:main-survey:Q3}) and after (\ref{app:main-survey:Q24}) exposure to My Activity.
We find that concern significantly decreases ($W=\num{2519.5}$, $p<0.001$) with an effect size of $r=\num{0.32}$, suggesting that this decrease in concern is moderate in size.

\begin{table}[t]
\centering
\caption[Post-exposure concern regression analysis]{\label{tab:post-concern-regression}
Ordinal regression model to describe the level of concern after visiting the My Activity dashboard.
The model uses a descending concern scale (\ie, from \emph{extremely} to \emph{not at all concerned}).
The Aldrich-Nelson pseudo $R^2$ of the model is \num[round-mode=places]{0.631858309728864}.
}
{
\small
\begin{tabular*}{\columnwidth}{
@{}
E{8.08926pt} 
@{\extracolsep{\fill}}
S[table-format = -2.2]
S[table-format = 3.2, scientific-notation = fixed, round-mode = places, round-minimum = 0.01, round-integer-to-decimal, tight-spacing = true] 
@{\extracolsep{0pt}}H@{\extracolsep{\fill}}
@{\extracolsep{0pt}}H@{\extracolsep{\fill}}
S[table-format = <1.3, scientific-notation = fixed, round-mode = places, round-precision = 3, round-minimum = 0.001]
@{\extracolsep{1pt}}
l
}
    \toprule
    {\textbf{Factor}} & {\textbf{Est.}} & {\textbf{OR}} & {\textbf{Error}} & {\textbf{t value}} & {\textbf{Pr(\textgreater\textbar z\textbar)}} & \\ 
    \midrule
    $\text{Pre-Exp. concern} = \textit{Extremely}$ & 5.71 & 302.8410699 & 1.03 & 5.56 & 2.74E-08 & *** \\ 
    $\text{Pre-Exp. concern} = \textit{Moderately}$ & 4.56 &  95.6011912 & 0.94 & 4.87 & 1.09E-06 & *** \\ 
    $\text{Pre-Exp. concern} = \textit{Somewhat}$ & 2.77 &  15.9609651 & 0.88 & 3.14 & 1.69E-03 & ** \\ 
    $\text{Pre-Exp. concern} = \textit{Slightly}$ & 1.18 &   3.2585878 & 0.86 & 1.38 & 1.67E-01 &  \\ 
    Increasing benefit & -0.17 &   0.8475816 & 0.37 & -0.45 & 6.54E-01 &  \\ 
    $\text{Knows My Activity} = Yes$ & -0.32 &   0.7247347 & 0.34 & -0.94 & 3.48E-01 &  \\ 
    $\text{IUIPC cont.} > 3.5$ & 0.29 &   1.3422555 & 1.07 & 0.27 & 7.84E-01 &  \\ 
    $\text{IUIPC awar.} > 3.5$ & -0.41 &   0.6607805 & 2.07 & -0.20 & 8.42E-01 &  \\ 
    $\text{IUIPC coll.} > 3.5$ & 0.29 &   1.3317022 & 0.54 & 0.53 & 5.95E-01 &  \\ 
    $\text{Gender} = Male$ & -0.24 &   0.7865547 & 0.34 & -0.71 & 4.81E-01 &  \\ 
    $\text{Age} \in\{18-34,~25-34\}$ & 0.39 &   1.4707258 & 0.48 & 0.81 & 4.17E-01 &  \\ 
    $\text{Age} \in\{35-44,~45-54\}$ & 0.91 &   2.4956446 & 0.47 & 1.95 & 5.15E-02 & $\cdot$ \\ 
    $\text{Edu.} \in \{\textit{No sch.g, (Sm.) HS}\}$ & 0.19 &   1.2054607 & 0.55 & 0.34 & 7.34E-01 &  \\ 
    $\text{Edu.} \in \{\textit{Sm. col., Assoc., Prof.}\}$ & 0.07 &   1.0735471 & 0.36 & 0.20 & 8.44E-01 &  \\ 
    $\text{Has IT background}$ & 0.64 &   1.9011462 & 0.40 & 1.62 & 1.05E-01 &  \\ 
    $\text{\# of activities} > \text{median}$ & -0.47 &   0.6269355 & 0.32 & -1.45 & 1.46E-01 &  \\ 
    \midrule
    \textbf{Intercepts} & & & & & \\
    \midrule
    $\textit{Not at all}~|~\textit{Slightly}$ & 0.58 &   1.7861953 & 2.01 & 0.29 & 7.73E-01 &  \\ 
    $\textit{Slightly}~|~\textit{Somewhat}$ & 2.87 &  17.6636336 & 2.03 & 1.41 & 1.58E-01 &  \\ 
    $\textit{Somewhat}~|~\textit{Moderately}$ & 4.63 & 102.0066655 & 2.05 & 2.26 & 2.40E-02 & * \\ 
    $\textit{Moderately}~|~\textit{Extremely}$ & 6.77 & 875.0841955 & 2.07 & 3.27 & 1.09E-03 & ** \\ 
    \bottomrule
\end{tabular*}
}
\footnotesize
\textbf{Signif. codes:} $\text{***}~\widehat{=} < 0.001$; $\text{**}~\widehat{=} <0.01$; $\text{*}~\widehat{=} <0.05$; $\cdot~\widehat{=} <0.1$
\end{table}

{}

\begin{table}[t]
\centering
\caption[Post-exposure frequency of benefit regression analysis]{\label{tab:post-benefit-regression}
Ordinal regression model to describe the frequency of benefit after after visiting the My Activity dashboard.
In the model a ascending frequency scale (\ie, from \emph{never} to \emph{always}) is used.
The Aldrich-Nelson pseudo $R^2$ of the model is \num[round-mode=places]{0.681841649836333}.
}
{
\small
\newsavebox{\equbox}
\begin{tabular*}{\columnwidth}{
@{}
E{8.08926pt} 
@{\extracolsep{\fill}}
S[table-format = -2.2]
S[table-format = 1.2e+1, tight-spacing = true] 
@{\extracolsep{0pt}}H@{\extracolsep{\fill}}
@{\extracolsep{0pt}}H@{\extracolsep{\fill}}
S[table-format = <1.3, scientific-notation = fixed, round-mode = places, round-precision = 3, round-minimum = 0.001]
@{\extracolsep{1pt}}
l
}
    \toprule
    {\textbf{Factor}} & {\textbf{Est.}} & {\textbf{OR}} & {\textbf{Error}} & {\textbf{t value}} & {\textbf{Pr(\textgreater\textbar z\textbar)}} & \\ 
    \midrule
    $\text{Pre-Exp. benefit} = \textit{Never}$ & 22.29 & 4.81E+09 & 0.70 & 31.72 & 7.42E-221 & *** \\ 
    $\text{Pre-Exp. benefit} = \textit{Rarely}$ & 20.26 & 6.28E+08 & 0.51 & 39.44 & 2.22E-308 & *** \\ 
    $\text{Pre-Exp. benefit} = \textit{Sometimes}$ & 18.58 & 1.17E+08 & 0.52 & 35.95 & 5.37E-283 & *** \\ 
    $\text{Pre-Exp. benefit} = \textit{Often}$ & 16.12 & 1.00E+07 & 0.66 & 24.45 & 4.49E-132 & *** \\ 
    Increasing concern & 0.57 & 1.77E+00 & 0.51 & 1.11 & 2.68E-01 &  \\ 
    $\text{Knows My Activity} = Yes$ & -0.56 & 5.71E-01 & 0.37 & -1.50 & 1.33E-01 & \\  
    $\text{IUIPC cont.} > 3.5$ & 0.00 & 1.00E+00 & 1.02 & 0.00 & 9.98E-01 &  \\ 
    $\text{IUIPC awar.} > 3.5$ & 0.18 & 1.20E+00 & 2.24 & 0.08 & 9.35E-01 &  \\ 
    $\text{IUIPC coll.} > 3.5$ & 0.41 & 1.50E+00 & 0.57 & 0.72 & 4.73E-01 &  \\ 
    $\text{Gender} = Male$ & 0.92 & 2.51E+00 & 0.37 & 2.46 & 1.38E-02 & * \\ 
    $\text{Age} \in\{18-34,~25-34\}$ & 0.76 & 2.14E+00 & 0.50 & 1.53 & 1.26E-01 &  \\ 
    $\text{Age} \in\{35-44,~45-54\}$ & 0.58 & 1.79E+00 & 0.51 & 1.13 & 2.56E-01 &  \\ 
    $\text{Edu.} \in \{\textit{No sch.g, (Sm.) HS}\}$ & -0.16 & 8.50E-01 & 0.59 & -0.28 & 7.82E-01 &  \\ 
    $\text{Edu.} \in \{\textit{Sm. col., Assoc., Prof.}\}$ & -0.16 & 8.51E-01 & 0.38 & -0.42 & 6.71E-01 &  \\ 
    $\text{Has IT background}$ & -0.07 & 9.30E-01 & 0.42 & -0.17 & 8.64E-01 &  \\ 
    $\text{\# of activities} > \text{median}$ & 0.30 & 1.36E+00 & 0.34 & 0.89 & 3.73E-01 &  \\ 
    \midrule
    \textbf{Intercepts} & & & & & \\
    \midrule
    $\textit{Always}~|~\textit{Often}$ & 15.61 & 6.00E+06 & 1.68 & 9.28 & 1.69E-20 & *** \\ 
    $\textit{Often}~|~\textit{Sometimes}$ & 18.78 & 1.43E+08 & 1.63 & 11.52 & 1.05E-30 & *** \\ 
    $\textit{Sometimes}~|~\textit{Rarely}$ & 21.99 & 3.55E+09 & 1.66 & 13.27 & 3.32E-40 & *** \\ 
    $\textit{Rarely}~|~\textit{Never}$ & 24.56 & 4.63E+10 & 1.71 & 14.34 & 1.15E-46 & *** \\ 
    \bottomrule
\end{tabular*}
}
\footnotesize
\textbf{Signif. codes:} $\text{***}~\widehat{=} < 0.001$; $\text{**}~\widehat{=} <0.01$; $\text{*}~\widehat{=} <0.05$; $\cdot~\widehat{=} <0.1$

\end{table}

{}

To explore what factors may have influenced the decline in concern, we performed ordinal logistic regression with outcome variable of the Likert concern scale (see \autoref{tab:post-concern-regression}).
We included binary variables for initial concern, benefit increased, high IUIPC factors, gender, age, education level, IT background, and number of activities stored in the Google account, and the final model had a Aldrich-Nelson pseudo-$R^2=0.63$ (see \autoref{tab:post-concern-regression-full} in \appref{app:regression-tables} for the complete model).
We find that those who had \emph{extremely} ($\eta=5.71,OR=303,p<0.001$), \emph{moderately} ($\eta=4.56,OR=96,p<0.001$), and \emph{somewhat} ($\eta=2.77,OR=16,p<0.001$) concern initially were significantly likely to reduce their concern after exposure.
Participants who were \emph{extremely} concerned were $303\times$ more likely to reduce their concern, and those \emph{moderately} concerned were $96\times$ more likely.
All other factors seem to have no or little effect, except perhaps for the age range \numrange{35}{54} ($\eta=0.91,OR=2.5,p=0.052$).

The alluvium plot in \autoref{fig:collectingConcern-alluvium} shows in more detail how the level of concern changes among the participants based on their initial concern.
In total, \num{61} (\SI{39.87}{\percent}) participants moved down the scale, \num{69} (\SI{45.10}{\percent}) stayed the same, and only \num{23} (\SI{15.03}{\percent}) increased their concern.

\paragraph{Changing Perceptions of Benefits.}
We find that there is a significant increase in perceived benefit (Wilcoxon signed-rank test, $W=\num{435}$, $p<0.001$) with a moderate effect ($r=\num{0.32}$).
Using the same factors as before, we constructed an ordinal logistic regression model to identify potential covariants that led to the increase in benefit (see \autoref{tab:post-benefit-regression}).
The full model is described in \autoref{tab:post-benefit-regression-full} in \appref{app:regression-tables}.

Across all initial benefit responses (\emph{never}, \emph{rarely}, \emph{sometimes}, \emph{often}, and \emph{always}), the regression exposes significant likelihood of keeping the same benefit or increasing benefits with odds ratio $>10^6$, suggesting that participants across the spectrum recognized benefits to Google's data collection.
We observed that participants identified as male also were significantly more likely to increase their benefit perceptions ($\eta=0.92,OR=2.5,p=0.014$), but other factors were not meaningfully significant.

\autoref{fig:frequencyBenefit-alluvium} provides more insights into the broad increase in perceived benefit.
In total, \num{45} (\SI{29.41}{\percent}) increased their benefit response, \num{93} (\SI{60.78}{\percent}) kept it the same, and only \num{15} (\SI{9.80}{\percent}) decreased their perceived benefit.

\paragraph{Final Perceptions of Concerns.}
Post exposure, participants were also asked to explain their final concern (\ref{app:main-survey:Q26a})  choices.
Qualitative coding revealed that while the total number of participants describing a privacy concern dropped from \num{79} (\SI{51.63}{\percent}) to 72 (\SI{47.06}{\percent}).
The number of participants who described privacy concerns about the amount of information collected increased to \num{21} (\SI{13.73}{\percent}) from \num{15} (\SI{9.80}{\percent}).
For example, P22 said:
\begin{quote}
    \small
    \emph{
        I'd say I'm a little more concerned now about how much is being collected.
        Especially with one of the random activities shown in the survey being well over a year old.
    }
\end{quote}

Similarly, some participants, \num{25} (\SI{16}{\percent}) versus \num{22} (\SI{14}{\percent}), mentioned security concerns, and prevalent codes that increased included data misuse ($n = \num{20}$), \eg, \enquote{I'm worried about the misuse of the data and security of it} (P127), and personal data being released ($n = \num{14}$), \eg, \enquote{There is always a chance that personally identifiable information can somehow be leaked to the Internet at large} (P147).

We also observed a slight increased in participants describing that they were now unconcerned with the data collection, \num{25} (\SI{16}{\percent}) verses \num{21} (\SI{14}{\percent}).
For example, participant P13 said:
\begin{quote}
    \small
    \emph{
        I am not concerned.
        Nothing's ever gone wrong as a result of what they collect.
        I don't have things to hide.
        I imagine the data collection helps me.
    }
\end{quote}

\paragraph{Final Perceptions of Benefits.}
Explaining their final benefits from Google's data collection (\ref{app:main-survey:Q27a}), qualitative coding revealed an increase in the number of participants who described benefits of suggestions: \num{70} (\SI{45.75}{\percent}) versus \num{46} (\SI{30.06}{\percent}).
For example participant P20 said, \enquote{YouTube recommendations are tailored around my activity, so that's beneficial to me,} and participant P119 said, \enquote{Many of Google's services offer useful personalized suggestions based on my data.}
We also found an increase in the number of participants who said that they found that access to their activity history beneficial: \num{26} (\SI{16.99}{\percent}) versus \num{15} (\SI{9.80}{\percent}, \eg, \enquote{I can go back to a websites I have viewed about specific things if I need to, find that song I really want to hear that I listened to last week, and make it easier to get places I may be returning to} (P18).
Fewer participants said they received no benefit from the data collection: \num{19} (\SI{12.42}{\percent}) versus \num{26} (\SI{16.99}{\percent}), like P86 who said, \enquote{How and why would I benefit from it when I didn't even know they are collecting information about my activities.}

\subsection{\ref{RQ:change-behavior}: Behavioral Change}
To answer our third research question \ref{RQ:change-behavior}, we surveyed participants about their willingness to take action after they have learned about Google's data collection practices.
We asked three closed-ended (\ref{app:main-survey:Q23}, \ref{app:main-survey:Q24}, and \ref{app:main-survey:Q25}) and three open-ended questions (\ref{app:main-survey:Q23a}, \ref{app:main-survey:Q24a}, and \ref{app:main-survey:Q25a}) to gauge participants' intentions to take action or change their behavior as a result of the exposure to My Activity.
The results of the three closed-ended questions are summarized in \autoref{fig:settingsActivityBehaivor-bar}.

\paragraph{Change Account Settings.}
We asked the participants to indicate whether they plan to adjust some of the (privacy) settings after seeing their My Activity page (\ref{app:main-survey:Q23}).
The results were almost the same for \emph{yes} ($n = \num{57}$; \SI{37.3}{\percent}) and \emph{no} ($n = \num{56}$; \SI{36.6}{\percent}) while \SI{26.1}{\percent} ($n=\num{40}$) of the participants were \emph{unsure}.

We constructed a logistic regression model to identify factors that predict the outcome of being willing to change settings.
We included covariants for change in concern, change in benefit, IUIPC responses, demographics, and total number of activities over the lifetime of an account.
The model did not expose any significant factors (see \appref{app:regression-tables} \autoref{tab:change-setting-regression-full}).

We also qualitatively explored participants views about their privacy settings by asking them which settings, if any, would they change (\ref{app:main-survey:Q23a}).
More participants ($n = \num{75}$; \SI{49}{\percent}) in their qualitative answer responded with a privacy setting that they would change.
We recognize that many of these participants do not plan to actually change settings given their prior quantitative responses.
We found that changing delete frequency ($n = \num{16}$), or stopping data collection ($n = \num{17}$), or changing how long information is stored for specific things ($n = \num{27}$) are the most popular reasons to revise the settings.
For example P22 said, \enquote{I would update when they delete my data so it stays current, relevant, and up-to-date.}
P75 said, \enquote{I would have my settings changed so that it no longer stores any data,} and P120 said, \enquote{Probably auto delete since I don't remember to go delete it often enough.}

Other participants ($n = \num{52}$; \SI{34}{\percent}) reported that they would not change their settings.
One reason for not changing the settings was that the participant likes the current settings.
For instance P13 said, \enquote{I have no complaints so see no reason to fix something that isn't broken.}
Another reason was that they had already configured the settings.
For example P133 said, \enquote{I've already used this page and configured it the way I want.}

Participants who were undecided ($n = \num{21}$; \SI{14}{\percent}) about changing the settings stated they wanted to review the settings.
Like participant P45 who said, \enquote{I need to look more into the settings to see something I may change.} Undecided participants also reported that they wanted to review the data collection, \eg, \enquote{I'd at least want to actually take a look and see just how much is collected, with using my account across all devices, and how far back my activity goes} (P22).

\figSettingsActivityBehaivorBar{}

\paragraph{Review or Delete Activities.}
When asked whether they plan to use My Activity again after the survey for reviewing or deleting activities (\ref{app:main-survey:Q24}), only \SI{36.6}{\percent} of the participants responded with \emph{yes}.
The remaining \SI{63.4}{\percent} ($n = \num{56}$) were either \emph{unsure} ($n = \num{48}$; \SI{31.4}{\percent}) or said \emph{no} ($n=\num{49}$; \SI{32.03}{\percent}).

We performed logistic regression to determine factors that would lead to reviewing activities (see \autoref{tab:review-activity-regression}; full model \autoref{tab:review-activity-regression-full}).
We found a significant correlation with IUIPC collection scale questions ($\beta=1.43, OR=4.19, p=0.042$), where participants with high privacy concerns regarding data collection were $4.19\times$ more likely to review activities later.
This finding suggests that individuals predisposed to have concerns about data collection are likely to benefit the most from My Activity.

We also qualitatively coded participants' explanations for why they would or would not review their activities (\ref{app:main-survey:Q24a}).
The main reason participants gave for continuing to use My Activity was to delete activities ($n = \num{90}$; \SI{58.8}{\percent}).
The most common activities participants said they would delete were Search ($n = \num{31}$), Maps ($n = \num{23}$), and YouTube ($n = \num{19}$).
For others ($n = \num{12}$) it was activities of a sensitive nature that they would return to delete.
For example, participant P89 said:
\begin{quote}
    \small
    \emph{
        Personal activities.
        Like I noticed that there were medical searches in my activities.
        It makes me uncomfortable that information is taken about me.
    }
\end{quote}

Other participants ($n = \num{41}$; \SI{26.8}{\percent}) reported that they do not plan to use My Activity in the future.
Reasons included making changes would be too time consuming ($n = \num{7}$), \eg, \enquote{I have better things to do with my time, frankly, than to be reviewing this} (P92), or that they would easily forget to do so ($n = \num{3}$), \eg, \enquote{Honestly, I'll probably forget about it, so I'm unlikely to delete things a month from now} (P129).

Still others were ($n = \num{12}$; \SI{7.8}{\percent}) undecided.
For instance participant P36 said, \enquote{I'm not sure, I would have to weigh convenience for me vs. the feeling of too much information being collected.}

\begin{table}[tbp]
\centering
\caption[Review activities regression analysis]{\label{tab:review-activity-regression}
Binomial logistic model to describe which other factors (beside visiting My Activity) influenced the participants plan to review/delete activities (\emph{yes} responses to Question \ref{app:main-survey:Q24}).
The  Aldrich-Nelson pseudo  $R^2 = \num[round-mode=places]{0.25321311230276}$.
}
{
\small
\begin{tabular*}{\columnwidth}{
@{}
E{8.08926pt} 
@{\extracolsep{\fill}}
S[table-format = -2.2]
S[table-format = 1.2e+1, tight-spacing = true]
@{\extracolsep{0pt}}H@{\extracolsep{\fill}}
@{\extracolsep{0pt}}H@{\extracolsep{\fill}}
S[table-format = <1.3, scientific-notation = fixed, round-mode = places, round-precision = 3, round-minimum = 0.001]
@{\extracolsep{1pt}}
l
}
    \toprule
    {\textbf{Factor}} & {\textbf{Est.}} & {\textbf{OR}} & {\textbf{Error}} & {\textbf{z value}} & {\textbf{Pr(\textgreater\textbar z\textbar)}} & \\ 
    \midrule
    (Intercept) & 13.77 & 9.53E+05 & 882.74 & 0.02 & 9.88E-01 &  \\ 
    Increasing concern & 0.90 & 2.45E+00 & 0.49 & 1.82 & 6.80E-02 & $\cdot$ \\ 
    Increasing benefit & 0.05 & 1.05E+00 & 0.41 & 0.13 & 9.00E-01 &  \\ 
    $\text{IUIPC cont.} > 3.5$ & 0.63 & 1.88E+00 & 1.25 & 0.51 & 6.13E-01 &  \\ 
    $\text{IUIPC awar.} > 3.5$ & -16.69 & 5.63E-08 & 882.74 & -0.02 & 9.85E-01 &  \\ 
    $\text{IUIPC coll.} > 3.5$ & 1.43 & 4.19E+00 & 0.71 & 2.03 & 4.24E-02 & * \\ 
    $\text{Gender} = Male$ & 0.45 & 1.56E+00 & 0.39 & 1.14 & 2.55E-01 &  \\ 
    $\text{Age} \in\{18-34,~25-34\}$ & -0.07 & 9.32E-01 & 0.53 & -0.13 & 8.93E-01 &  \\ 
    $\text{Age} \in\{35-44,~45-54\}$ & 0.24 & 1.27E+00 & 0.53 & 0.45 & 6.52E-01 &  \\ 
    $\text{Edu.} \in \{\textit{No sch.g, (Sm.) HS}\}$ & 0.85 & 2.34E+00 & 0.63 & 1.34 & 1.79E-01 &  \\ 
    $\text{Edu.} \in \{\textit{Sm. col., Assoc., Prof.}\}$ & -0.04 & 9.64E-01 & 0.41 & -0.09 & 9.28E-01 &  \\ 
    $\text{Has IT background}$ & 0.34 & 1.40E+00 & 0.44 & 0.77 & 4.41E-01 &  \\ 
    $\text{\# of activities} > \text{median}$ & -0.35 & 7.02E-01 & 0.36 & -0.97 & 3.30E-01 &  \\ 
    \bottomrule
\end{tabular*}
}
\footnotesize
\textbf{Signif. codes:} $\text{***}~\widehat{=} < 0.001$; $\text{**}~\widehat{=} <0.01$; $\text{*}~\widehat{=} <0.05$; $\cdot~\widehat{=} <0.1$
\end{table}

{}

\begin{table}[t]
\centering
\caption[Use Google differently regression analysis]{\label{tab:google-different-regression}
Binomial logistic model to describe which other factors 
influenced the participants plan to use Google products differently in the future (\emph{yes} responses to Question \ref{app:main-survey:Q25}).
The Aldrich-Nelson pseudo $R^2=\num[round-mode=places]{0.420131390603509}$.
}
{
\small
\begin{tabular*}{\columnwidth}{
@{}
E{8.08926pt} 
@{\extracolsep{\fill}}
S[table-format = -2.2]
S[table-format = 2.2e+1, tight-spacing = true]
@{\extracolsep{0pt}}H@{\extracolsep{\fill}}
@{\extracolsep{0pt}}H@{\extracolsep{\fill}}
S[table-format = <1.3, scientific-notation = fixed, round-mode = places, round-precision = 3, round-minimum = 0.001]
@{\extracolsep{1pt}}
l
}   
    \toprule
    {\textbf{Factor}} & {\textbf{Est.}} & {\textbf{OR}} & {\textbf{Error}} & {\textbf{z value}} & {\textbf{Pr(\textgreater\textbar z\textbar)}} & \\ 
    \midrule
    (Intercept) & -16.23 & 8.93E-08 & 3956.18 & -0.00 & 9.97E-01 &  \\ 
    Increasing concern & 2.50 & 1.22E+01 & 0.62 & 4.01 & 6.04E-05 & *** \\ 
    Increasing benefit & -1.31 & 2.71E-01 & 0.59 & -2.21 & 2.73E-02 & * \\ 
    $\text{IUIPC cont.} > 3.5$ & 18.42 & 1.00E+08 & 1361.72 & 0.01 & 9.89E-01 &  \\ 
    $\text{IUIPC awar.} > 3.5$ & -4.82 & 8.06E-03 & 4183.97 & -0.00 & 9.99E-01 &  \\ 
    $\text{IUIPC coll.} > 3.5$ & 1.40 & 4.07E+00 & 0.88 & 1.60 & 1.09E-01 &  \\ 
    $\text{Gender} = Male$ & -0.73 & 4.80E-01 & 0.51 & -1.45 & 1.46E-01 &  \\ 
    $\text{Age} \in\{18-34,~25-34\}$ & 0.97 & 2.63E+00 & 0.75 & 1.29 & 1.98E-01 &  \\ 
    $\text{Age} \in\{35-44,~45-54\}$ & 1.21 & 3.36E+00 & 0.76 & 1.60 & 1.09E-01 &  \\ 
    $\text{Edu.} \in \{\textit{No sch.g, (sm) HS}\}$ & 0.45 & 1.57E+00 & 0.78 & 0.58 & 5.62E-01 &  \\ 
    $\text{Edu.} \in \{\textit{Sm col. Assoc., Prof.}\}$ & -0.16 & 8.56E-01 & 0.50 & -0.31 & 7.57E-01 &  \\ 
    $\text{Has IT background}$ & 0.86 & 2.37E+00 & 0.55 & 1.58 & 1.14E-01 &  \\ 
    $\text{\# of activities} > \text{median}$ & -1.25 & 2.86E-01 & 0.46 & -2.72 & 6.49E-03 & ** \\ 
    \bottomrule
\end{tabular*}
}
\footnotesize
\textbf{Signif. codes:} $\text{***}~\widehat{=} < 0.001$; $\text{**}~\widehat{=} <0.01$; $\text{*}~\widehat{=} <0.05$; $\cdot~\widehat{=} <0.1$
\end{table}

{}

\paragraph{Use Google Differently.}
Nearly \SI{50}{\percent} ($n = \num{76}$) of the participants did not plan to use Google products differently in the future in response to \ref{app:main-survey:Q24}.
The remaining responses  split evenly between \emph{yes} ($n = \num{39}$; \SI{25.5}{\percent}) and \emph{unsure} ($n = \num{38}$; \SI{24.8}{\percent}).

We performed a logistic regression to determine factors that may influence reported changes in behavior (see \autoref{tab:google-different-regression}; full model \autoref{tab:google-different-regression-full}).
Unsurprisingly, we found two significant factors.
Those who had an increase in concern ($\beta=2.50,OR=12.2,p<0.001$) and a decrease in (or same) benefit ($\beta=-1.31,OR=0.27,p=0.027$) were significantly more likely to use Google products differently.
This represents a small fraction of participants in our study: 23 (\SI{15.03}{\percent}) participants noted an increase in concern, 15 (\SI{9.80}{\percent}) reported a decrease in benefit.

In addition, we found a third significant factor.
Participants whose accounts contained a high number of activities (\ie, more than the median number of activities) were significantly more likely to report to use Google products differently ($\beta=-1.25,OR=0.29,p=0.006$).

Looking at the qualitative results shows that of those who planned to use Google products differently some would change settings ($n = \num{14}$; \SI{9}{\percent}), such as limiting data collection or deleting their activities more often, \eg, \enquote{I am definitely going to be turning off history for YouTube while working} (P147).
Others would change the way they use Google products and services more generally ($n = \num{12}$; \SI{7.8}{\percent}), such as being more careful when using them, \eg, \enquote{I'd certainly be aware of what was being collected and modify my searches accordingly} (P148).
Some participants would start to limit their usage of Google products and services ($n = \num{9}$; \SI{5.9}{\percent}), \eg, \enquote{I would use less of Google and more of other services} (P96).

Of those participants who were unsure if they would change using Google products, some ($n = \num{10}$; \SI{6.5}{\percent}) stated that change would be difficult because of the importance of Google products; for instance P6 said:
\begin{quote}
    \small
    \emph{
        I realize Google products are necessary to my lifestyle and work, but I also like to be in control of my data.
        I'm not sure what the best course of action is at this point.
    }
\end{quote}

Of the participants who would not change the way they use Google products, many ($n = \num{37}$; \SI{24}{\percent}) claimed they were happy with the status quo, like participant P139 who said, \enquote{I am happy with the current setup and will continue as I always have.}
Some participants ($n = \num{25}$; \SI{16}{\percent}) were simply unconcerned, \eg, \enquote{I just don't care enough from what I saw to change how I use Google} (P122).
Others ($n = \num{6}$; \SI{3.9}{\percent}) are simply privacy resigned, \eg, \enquote{I've accepted the fact that they work this way whether I view it as right or not} (P120).

\paragraph{Willingness to Pay for Google's Services.}
We asked participants if they were willing to pay for Google services if activity data were not collected (\ref{app:main-survey:Q16}), and those results are presented in \autoref{fig:paidPlan-bar}.
Nearly half of the responses ($n = \num{74}$; \SI{47.77}{\percent}) would not pay, which is in line with previous work~\cite{chanchary-15-percep-tracking}, but 70 (\SI{45.75}{\percent}) say they would pay at least \SI{1}[\dollar]{\USD} per-month, with large clusters at \SI{10}[\dollar]{\USD}, \SI{5}[\dollar]{\USD}, and \SI{1}[\dollar]{\USD} per-month.
Only 10 (\SI{6.5}{\percent}) described a willingness to pay more than \SI{10}[\dollar]{\USD} per month.

The average revenue per user~(ARPU) is currently not reported by Google and differs significantly between regions.
According to eMarketer~\cite{clark-20-google-us-net-ad-revenue} Google will make a \emph{net ad revenue} (after paying traffic acquisition costs to partner sites) in the U.S. of approximately \SI{39.58}[\dollar]{\billion\USD} in 2020.
As of December 2020, Google had close to 271 million unique monthly visitors in the U.S.~\cite{comscore-20-unique-us-visitors}, resulting in an ARPU of $\sim \SI{146}[\dollar]{\USD}$ (Facebook~\SI{159}[\dollar]{\USD}~\cite{facebook-21-earnings}), or roughly \SI{12}[\dollar]{\USD} per month.
This is in line with Google's pricing for workspace accounts (\SI{12}[\dollar]{\USD}  per-user and per-month), and thus, one can assume that Google would require a fee of $\geq \SI{12}[\dollar]{\USD}$  per month (but likely more) in return for not collecting data.
Our data suggests that only a small fraction would pay such a fee, and perhaps fewer, as this result could be affected by response bias.

\figPaidPlanBar{}

\section{Discussion and Conclusion}\label{sec:discussion}
\paragraph{Controlling Data Collection.}
Participants reported significantly higher benefits from and lower concerns about Google's data collection after interacting with My Activity.
These shifts could be accounted for by an increased awareness of the collection and the prospects of being in control of that collection with access to the history of activities, similar to what Schnorf \etal found~\cite{schnorf-14-transparency-trust}.
This may especially be the case in our study as only $\sim\SI{50}{\percent}$ of respondents reported being aware of specific activity displayed during the survey, and $\sim\SI{75}{\percent}$ agreed or strongly-agreed that My Activity helps them to better understand Google's data collection.

The notion of information flow controls is an important factor in privacy perception~\cite{tavani-07-theories-of-privacy}.
Interactions with My Activity increase the subjective (and also objective) control over collected data, reducing concern in relation to an original feelings of lack of control and an inability to restrict data access.
My Activity and other data collection transparency and management tools are both in the end-users' and service providers' best interest, and we expect (and hope) that more online services will provide such tools in the future.

\paragraph{Opaque Control Choices.}
My Activity allows users a plethora of choices, but it may be too difficult for users to make informed decisions about individual activities as the impact of keeping or deleting individual items is opaque to the user.
In their study of ad profiles, Rao \etal suggested that the dashboards did not provide transparency on how or why user profiles were created~\cite{rao-14-what-do-they-know}, and this lack of additional information inhibits clear decision making.
The only current explanations on My Activity suggest that the experience will degrade, but specifically why deleting any given activity or bundle of activities degrades experiences (or improves privacy) is not readily available.

There is evidence that providing some more insight into inferences could be beneficial, as users tend to relate inferences with their past activities~\cite{rader-20-inferences-vs-perceptions}, and there already exists language from Google that suggests the seasonality of data matters in inferences; this motivates the deletion time frames of 3- and 18-months~\cite{pierce-20-google-3-and-18-months}.
Expanding the options for how to manage data collection, perhaps based on inferences made or other metrics, would better assist users in making clearer choices in managing their activity data.

\paragraph{Management at Scale.}
Services like My Activity put significant pressure on users to continuously and individually manage their data collection, especially, as new data collection occurs all the time, and in some cases, users may have to review activities across multiple accounts.
It is likely that well intentioned data-privacy laws, like GDPR, may lead to increased data collection management due to data access requirements.
The truth is, such management does not scale, and the benefits of increased control could be neutered by the increase in the scale of decision making.

We see evidence for My Activity that users are unlikely to take advantage of these controls, perhaps due to the scale of the management requirement not just with Google but elsewhere.
This is only one of possibly many transparency tools; mechanisms for secure and transparent umbrella management of data collection across services is likely to be needed as privacy dashboards and data-rights laws proliferate.
Such umbrella services have been proposed previously in the literature~\cite{angulo-15-data-track, fischer-huebner-16-priv-dashboard, raschke-17-usable-pri-dashboard, weinshel-19-places-you-been}, and with apps like \emph{Jumbo}~\cite{newton-19-jumbo-app, miller-21-ios-tracking-transparency}, some first real-world tools exist.
However, these umbrella services need to find the balance between displaying relevant information to the users but not overwhelming them.
Exploring whether and how these services could reduce management at scale is an area of future research.

\paragraph{Lack of Negative Consequences.}
Only about a quarter of respondents indicated they would change their behavior as a result of increased awareness of Google's data collection, and this sentiment is entirely understandable given how integral Google's products and services are in the online experience.
A number of participants explicitly noted that they trust Google, and thus after exposure to their My Activity page remained unconcerned about Google's data collection.
As there is no information of potential privacy risks on the My Activity page, it is not surprising that people are unconcerned with and not aware of how their data can be used in expected ways by either Google or third-parties.

For those who indicated they would change settings, this group reported high on the IUIPC Control scale, indicating that individuals who seek more control over their data will likely take advantage of such a service.
The remaining users are less likely to do so, perhaps because they have not experienced negative consequence and instead rely on the default policy, which may not be in their best interest.
Increased exposure and encouragement for users to understand the benefits and risks of data collection could lead to better outcomes for everyone, as it may encourage service providers to use better default privacy settings.

\paragraph{Design Implications.}
Based on the findings of our study, we offer some suggestions to improve the utility of privacy dashboards.
\begin{enumerate*}[label=\arabic*)]
    \item Provide concrete explanations for which purpose activities are collected and stored.
    For instance, when activities are used to infer interests of a person, make this link between the activity and inference more explicit (\eg, search query for \enquote{Seattle Seahawks} results in the inference \enquote{American Football} and the aggregated inference \enquote{Sports}).
    
    \item Participants felt overwhelmed with the amount of activities being collected and presented to them.
    It is worthwhile to explore ways to give users a better overview of and means to navigate through their  activities.
    Showing simple statistics (\eg, the number of activities grouped by month or service) might helping people to better grasp the amount of activities collected.
    Activities could also be further clustered beyond the existing My Activity bundle view, which groups by time and Google product.
    For example, each cluster could be further grouped by broader themes, e.g., by inferences applied for advertising, that could assists users to better focus on activities that may need manual review.
    
    \item Some participants expressed the need for better ways to remove certain activity classes, for example, any search related to medical issues.
    Offering keyword management strategies where users can custom define activity deletion policies based on user defined criteria would help users manage their privacy without having to regularly inspect their activities.
    In May 2021, after data collection, Google introduced a ``quick delete'' feature which removes the last 15~minutes of search activities~\cite{fitzpatrick-21-delete-last-15-mins}.
\end{enumerate*}

However, adding too much functionality  carries the risk of overwhelming users with a complex UI, discouraging its use.
The simplicity of My Activity's design is admirable, but this needs to be balanced with providing substantive information about the purpose the data is collected and how it will be used.
Designing a more effective transparency tool that is both simple and deeply informative requires more exploration.

\section{Conclusion}\label{sec:conclusion}
In this work, we sought to understand how privacy dashboards and transparency tools affect concerns about and benefits from data collection.
Focusing on Google's My Activity tool, we conducted a pre-post-study where participants answered questions about concern/benefit before and after exposure to My Activity.
We find that My Activity significantly decreases concern about Google's data collection practices and increases the perceived benefit, despite participants qualitatively stating the same concerns and benefits before and after exposure.
Transparency tools, like My Activity, are clearly beneficial to the service providers and can also support data management for the user.
We, unfortunately, find that most participants are unwilling or unsure if they will review their activities following this study.

\section*{Acknowledgments}
We wish to thank Elissa Redmiles, Olivia Legault, and Genevieve Flynn. This research was partially funded by the MKW-NRW research training group SecHuman and the Deutsche Forschungsgemeinschaft (DFG, German Research Foundation) under Germany's Excellence Strategy  -- EXC 2092 CASA -- 390781972.

\clearpage
{
\hbadness 10000

\begin{footnotesize}
\bibliographystyle{plain}
\bibliography{main.bib}
\end{footnotesize}
}

\clearpage
\appendix
\vspace{-.1in}
\section{Appendix}\label{sec:appendix}

\subsection{Screening Survey Instrument}\label{app:pre-survey}
\setlength{\columnsep}{0pt}
\setlength{\multicolsep}{0pt}
\definecolor{structure}{HTML}{03588C} 
\definecolor{note}{HTML}{BF2C47} 

\begin{scriptsize}
Thank you for your interest in our survey. Your answers are important to us. \textbf{Please read the following instructions carefully:}
\begin{enumerate*}[label=(\roman*)]
    \item Take your time in reading and answering the questions.
    \item Answer the questions as accurately as possible.
    \item It is okay to say that you don’t know an answer.
\end{enumerate*}

\begin{questions}[label=\textbf{S\arabic*}]
    \item Do you have a personal Gmail address (an email address ending in ``gmail.com'')?
    \begin{multicols}{2}
        \begin{answers}
            \item Yes
            \item No
        \end{answers}
    \end{multicols}

    \item How long do you have that Gmail address?
    \begin{multicols}{2}
        \begin{answers}
            \item Less than a year
            \item One year
            \item Three years
            \item Five years
            \item More than five years
            \item I do not have a Gmail address
            \item Unsure
        \end{answers}
    \end{multicols}
    
    \item\label{app:pre-survey:S3} Which other Google products do you currently use? (Select all that apply.)
    \begin{multicols}{3}
        \begin{answers}
            \item Gmail
            \item Google Maps
            \item YoutTube
            \item Google Chrome
            \item Google Search
            \item Google Play
            \item Google Drive
            \item Google News
            \item Google Pay
            \item Android device
            \item None of these
        \end{answers}
    \end{multicols}
    
    \item[\textbf{A0}] Google began in January 1996 as a research project. Its initial public offering took place on August 19, 2004. Did the initial public offering of Google take place in 1996?
    \begin{multicols}{3}
        \begin{answers}
            \item Yes
            \item No
        \end{answers}
    \end{multicols}
    \begin{answers}
        \item Other (please specify) \hrulefill
    \end{answers}
    
    \item How frequently do you use these products?
    \textcolor{note}{\emph{[Included only products selected in \ref{app:pre-survey:S3}. If \enquote{None of these} was selected question was hidden]}}
    \begin{tabular*}{\linewidth}{@{}l@{\extracolsep{\fill}}*{6}{c@{}}}
        \toprule
                       & Always     & Often      & Sometimes  & Rarely     & Never      & Unsure     \\
        \midrule
        Gmail          & $\bigcirc$ & $\bigcirc$ & $\bigcirc$ & $\bigcirc$ & $\bigcirc$ & $\bigcirc$ \\
        Google Maps    & $\bigcirc$ & $\bigcirc$ & $\bigcirc$ & $\bigcirc$ & $\bigcirc$ & $\bigcirc$ \\
        YoutTube       & $\bigcirc$ & $\bigcirc$ & $\bigcirc$ & $\bigcirc$ & $\bigcirc$ & $\bigcirc$ \\
        Google Chrome  & $\bigcirc$ & $\bigcirc$ & $\bigcirc$ & $\bigcirc$ & $\bigcirc$ & $\bigcirc$ \\
        Google Search  & $\bigcirc$ & $\bigcirc$ & $\bigcirc$ & $\bigcirc$ & $\bigcirc$ & $\bigcirc$ \\
        Google Play    & $\bigcirc$ & $\bigcirc$ & $\bigcirc$ & $\bigcirc$ & $\bigcirc$ & $\bigcirc$ \\
        Google Drive   & $\bigcirc$ & $\bigcirc$ & $\bigcirc$ & $\bigcirc$ & $\bigcirc$ & $\bigcirc$ \\
        Google News    & $\bigcirc$ & $\bigcirc$ & $\bigcirc$ & $\bigcirc$ & $\bigcirc$ & $\bigcirc$ \\
        Google Pay     & $\bigcirc$ & $\bigcirc$ & $\bigcirc$ & $\bigcirc$ & $\bigcirc$ & $\bigcirc$ \\
        Android device & $\bigcirc$ & $\bigcirc$ & $\bigcirc$ & $\bigcirc$ & $\bigcirc$ & $\bigcirc$ \\
        \bottomrule
    \end{tabular*}
    
    \item How important is using Google products to your Internet experience?
    \begin{multicols}{2}
        \begin{answers}
            \item Not important
            \item Slightly important
            \item Moderately important
            \item Important
            \item Very important
        \end{answers}
    \end{multicols}
    
    \item[] \textcolor{note}{\emph{[In the following, we asked the $10\times$ IUIPC items of the control, awareness, and collection scale as described by Malhotra \etal~\cite{malhotra-2004-iuipc}}.]}

\end{questions}

\end{scriptsize}

\subsection{Main Survey Instrument}\label{app:main-survey}

\begin{scriptsize}

\begin{questions}
    \item How aware are you of the amount of information that Google is collecting about your activities online?
    \label{app:main-survey:Q1}
    \begin{multicols}{2}
        \begin{answers}
            \item Not at all aware
            \item Slightly aware
            \item Somewhat aware
            \item Moderately aware
            \item Extremely aware
        \end{answers}
    \end{multicols}
    
    \item How concerned are you with the amount of information Google is collecting about your activities online?
    \label{app:main-survey:Q2}
    \begin{multicols}{2}
        \begin{answers}
            \item Not at all concerned
            \item Slightly concerned
            \item Somewhat concerned
            \item Moderately concerned
            \item Extremely concerned
        \end{answers}
    \end{multicols}
    \begin{questions}
        \item Please explain why.
        \label{app:main-survey:Q2a}
        \begin{answers}[widest=Answer:]
            \item[Answer:] \hrulefill
        \end{answers}
    \end{questions}
    
\pagebreak
    
    \item How often do you benefit from the amount of information that Google collects about your activities online?
    \label{app:main-survey:Q3}
    \begin{multicols}{2}
        \begin{answers}
            \item Never
            \item Rarely
            \item Sometimes
            \item Often
            \item Always
        \end{answers}
    \end{multicols}
    
    \begin{questions}
        \item Please explain why.
            \label{app:main-survey:Q3a}
        \begin{answers}[widest=Answer:]
            \item[Answer:] \hrulefill
        \end{answers}
    \end{questions}
\end{questions}

\noindent Some people use strategies to limit the amount of information that companies can collect about them online.
\begin{questions}[resume]
    \item Do you have any strategies for managing the kind of information Google may collect about you?
    \label{app:main-survey:Q4}
    \begin{answers}[widest=Answer:]
        \item[Answer:] \hrulefill
    \end{answers}
\end{questions}

\paragraph{What is Google My Activity?}\mbox{}\\
The following briefly introduces you to Google’s My Activity page.
For every account, Google provides an overview called My Activity, which contains the history of activities of your interactions with Google products.
Below is Google’s description of the My Activity page.
\begin{itemize}[leftmargin=*]
    \item What is My Activity?
    \begin{itemize}[leftmargin=*, nosep]
        \item \emph{\enquote{My Activity is a central place to view and manage activity like searches you've done, websites you've visited, and videos you've watched.}}\footnote{ \url{https://support.google.com/accounts/answer/7028918}, as of \today}
    \end{itemize}
    \item What are activities?
    \begin{itemize}[leftmargin=*, nosep]
        \item \emph{\enquote{When you use certain Google services, like Search, YouTube, or Chrome, your activity can be saved as data to your account.}}\addtocounter{footnote}{-1}\addtocounter{Hfootnote}{-1}\footnotemark 
    \end{itemize}
\end{itemize}

\paragraph{Google Login Page}\mbox{}\\
This survey requires that you login to your primary Google account for accessing items in your My Activity page.

\smallskip
\noindent\textbf{Privacy Note:}
We do not track or store your email address as part of this study, and we will not be able to tie your email address to any results or analysis. The researchers will never see your email address. At no time do the researchers have access to your Google account.


\paragraph{Explore My Activity}\mbox{}\\
In the next part of the study, we will ask you to explore Google’s My Activity page for your Google account.
You will have an opportunity to interact with your Google My Activity page for one minute and will then be returned to the survey.

\smallskip
\noindent\textbf{Privacy Note:}
We do not track or store your email address as part of this study, and we will not be able to tie your email address to any results or analysis. The researchers will never see your email address. At no time do the researchers have access to your Google account.


\begin{center}
\emph{Participants explored their My Activity page.}
\smallskip
\includegraphics[width=.95\columnwidth]{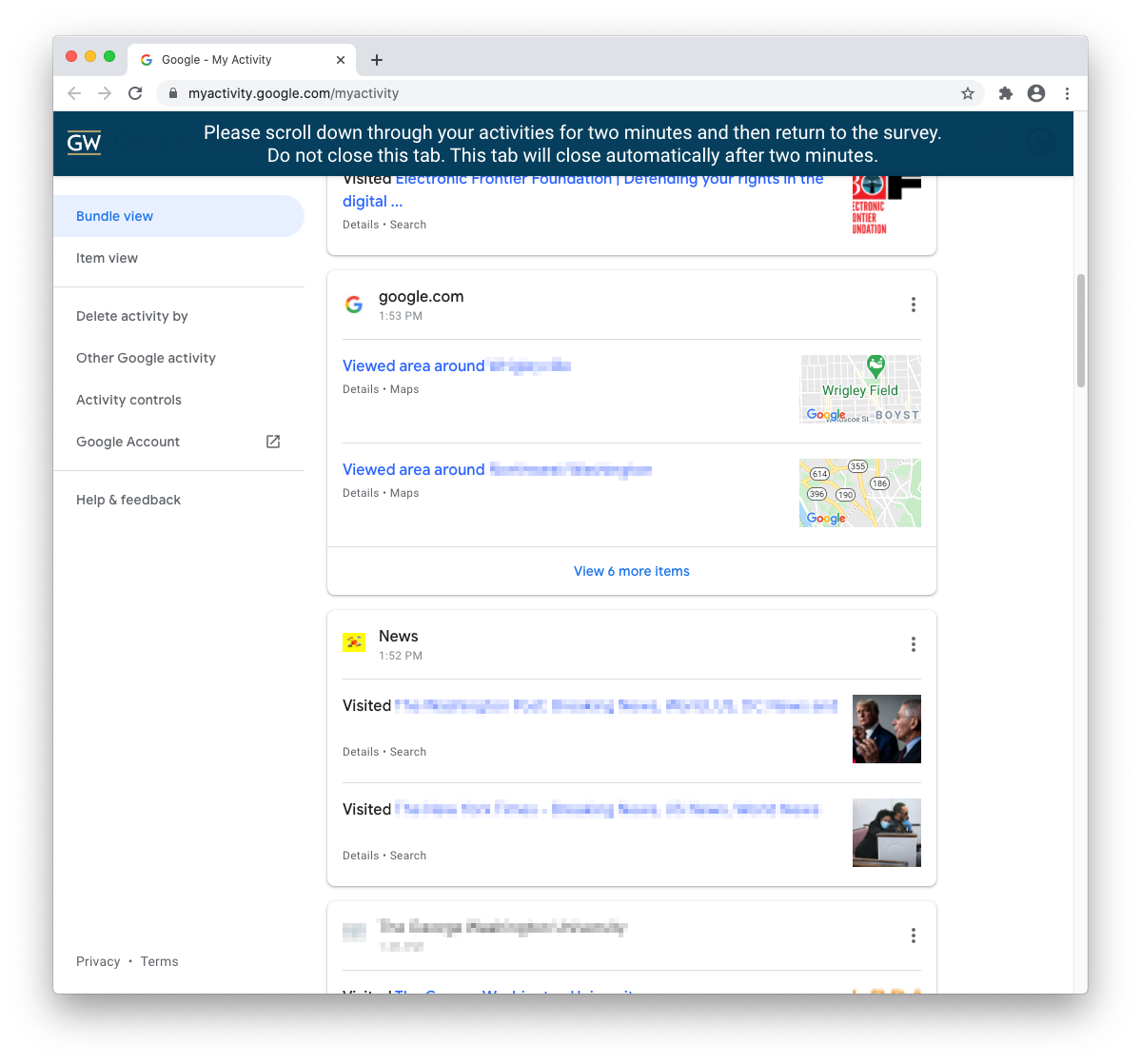}
\end{center}
\vspace{-1cm}

\begin{questions}[resume]
    \item Please provide any immediate reactions you have to exploring the My Activity page.
    \label{app:main-survey:Q5}
    \begin{answers}[widest=Answer:]
        \item[Answer:] \hrulefill
    \end{answers}
    
    \item Have you visited the My Activity page prior to this study?
    \label{app:main-survey:Q6}
    \begin{multicols}{3}
        \begin{answers}
            \item Yes
            \item No
            \item Unsure
        \end{answers}
    \end{multicols}
    
    \item[\textbf{A1}] What is the shape of a red ball?
    \begin{multicols}{2}
        \begin{answers}
            \item Red
            \item Round
            \item Blue
            \item Square
        \end{answers}
    \end{multicols}
    
    \item Provide three purposes for which you think Google is using your activity data.
    \label{app:main-survey:Q7}
    {
    \setlength{\columnsep}{.75pc}
    \begin{multicols}{3}
        \begin{enumerate}[leftmargin=*]
            \item \hrulefill
            \item \hrulefill
            \item \hrulefill
        \end{enumerate}
    \end{multicols}
    }

    \begin{questions}
        \item Based on your answer before, do you believe the purposes for Google using this information is beneficial to you in any way?
        \label{app:main-survey:Q7a}
        \begin{tabular*}{\linewidth}{@{}l@{\extracolsep{\fill}}*{5}{c}}
            \toprule
                       & \multicolumn{1}{@{}c@{}}{Not at all}
                       & \multicolumn{1}{@{}c@{}}{Slightly}
                       & \multicolumn{1}{@{}c@{}}{Somewhat}
                       & \multicolumn{1}{@{}c@{}}{Moderately}
                       & \multicolumn{1}{@{}c@{}}{Extremely}                            \\
            \cmidrule{2-6}
                       & \multicolumn{5}{@{}c@{}}{\emph{Beneficial}}                           \\
            \midrule
            1. Purpose & $\bigcirc$ & $\bigcirc$ & $\bigcirc$ & $\bigcirc$ & $\bigcirc$ \\
            2. Purpose & $\bigcirc$ & $\bigcirc$ & $\bigcirc$ & $\bigcirc$ & $\bigcirc$ \\
            3. Purpose & $\bigcirc$ & $\bigcirc$ & $\bigcirc$ & $\bigcirc$ & $\bigcirc$ \\
            \bottomrule
        \end{tabular*}
        
        \item Based on your answer before, do you believe the purposes for Google using this information is harmful to you in any way?
        \label{app:main-survey:Q7b}
        \begin{tabular*}{\linewidth}{@{}l@{\extracolsep{\fill}}*{5}{c}}
            \toprule
                       & \multicolumn{1}{@{}c@{}}{Not at all}
                       & \multicolumn{1}{@{}c@{}}{Slightly}
                       & \multicolumn{1}{@{}c@{}}{Somewhat}
                       & \multicolumn{1}{@{}c@{}}{Moderately}
                       & \multicolumn{1}{@{}c@{}}{Extremely}                            \\
            \cmidrule{2-6}
                       & \multicolumn{5}{@{}c@{}}{\emph{Harmful}}                              \\
            \midrule
            1. Purpose & $\bigcirc$ & $\bigcirc$ & $\bigcirc$ & $\bigcirc$ & $\bigcirc$ \\
            2. Purpose & $\bigcirc$ & $\bigcirc$ & $\bigcirc$ & $\bigcirc$ & $\bigcirc$ \\
            3. Purpose & $\bigcirc$ & $\bigcirc$ & $\bigcirc$ & $\bigcirc$ & $\bigcirc$ \\
            \bottomrule
        \end{tabular*}
    \end{questions}
    
    \item Are there any other concerns you might have with Google collecting this information?
    \label{app:main-survey:Q8}
    \begin{answers}[widest=Answer:]
        \item[Answer:] \hrulefill
    \end{answers}
  
    \item Do you believe your experience using Google services is improved by Google collecting this information?
    \label{app:main-survey:Q9}
    \begin{multicols}{2}
        \begin{answers}
            \item Strongly disagree
            \item Disagree
            \item Neither agree nor disagree
            \item Agree
            \item Strongly agree
        \end{answers}
    \end{multicols}
        
\end{questions}

\paragraph{Activity Presentation}\mbox{}\\
In the next part of the survey, we will ask you questions about \textbf{nine activities} from your My Activity page. The activities are chosen randomly. We do not collect information about that activity as part of this survey. That information remains private, \textbf{only accessible to you} and Google. We only note which service the activity is associated with, \eg, \enquote{Google search} vs. \enquote{YouTube view}, and the date on which it occurred. Further details are not collected as part of this survey.

\begin{center}
\includegraphics[width=0.8\columnwidth]{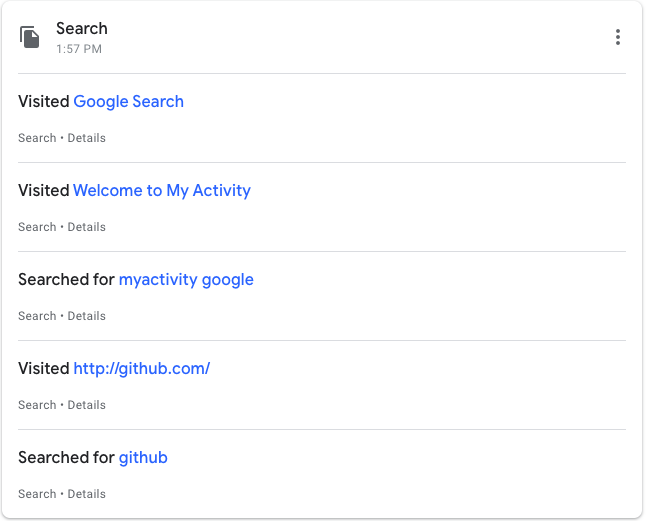}
\end{center}

\begin{center}
\includegraphics[width=0.8\columnwidth]{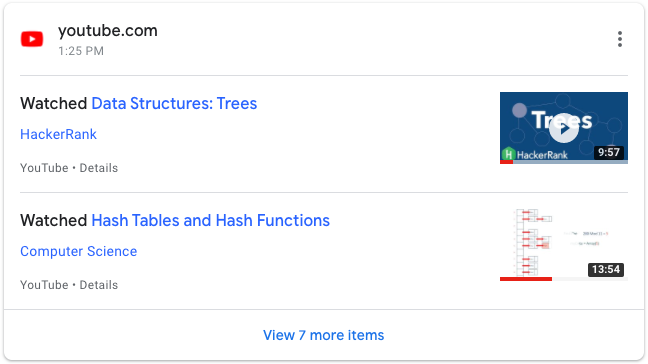}
\end{center}

\begin{center}
\includegraphics[width=0.8\columnwidth]{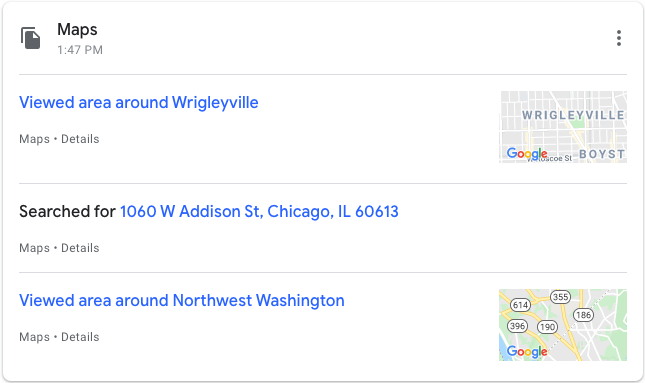}
\end{center}
    
\begin{questions}[resume]

    \item\label{app:main-survey:Q10} Do you recall this activity?
    \begin{multicols}{3}
        \begin{answers}
            \item Yes
            \item No
        \end{answers}
    \end{multicols}
    
    \item Prior to seeing this activity, have you been aware that Google stored this activity?
    \label{app:main-survey:Q11}
    \begin{multicols}{3}
        \begin{answers}
            \item Yes
            \item No
            \item Unsure
        \end{answers}
    \end{multicols}
    
    \item Storing this activity is necessary for my experience with using \textcolor{note}{\emph{[Google product name]}}.
    \label{app:main-survey:Q12}
    \begin{multicols}{2}
        \begin{answers}
            \item Strongly disagree
            \item Disagree
            \item Neither agree nor disagree
            \item Agree
            \item Strongly agree
        \end{answers}
    \end{multicols}
  
    \item Storing this activity changes my experience with using \textcolor{note}{\emph{[Google product name]}} in the following way:
    \label{app:main-survey:Q13}
        \begin{answers}
            \item Greatly harms my experience
            \item Harms my experience
            \item Slightly harms my experience
            \item Does not change my experience
            \item Slightly improves my experience
            \item Improves my experience
            \item Greatly improves my experience
        \end{answers}
    
    \item\label{app:main-survey:Q14} If you were to change how long this data is stored, when would you want it to be deleted?
        \begin{answers}
            \item Immediately, I do not want this data to be collected
            \item After a few hours
            \item After a day
            \item After a week
            \item After a month
            \item After 3 months
            \item After 18 months
            \item I wouldn't delete
        \end{answers}
        \textcolor{note}{\emph{[Repeat questions \ref{app:main-survey:Q10} to \ref{app:main-survey:Q14} for each activity presented to the participant]}}
\end{questions}

\begin{questions}[resume]
    \item Describe two feelings you had after viewing the activities we showed you.
    \label{app:main-survey:Q15}
    {
    \setlength{\columnsep}{.75pc}
    \begin{multicols}{2}
        \begin{enumerate}[leftmargin=*]
            \item \hrulefill
            \item \hrulefill
        \end{enumerate}
    \end{multicols}
    }
    
    \item\label{app:main-survey:Q16} If Google offered a paid plan where they do not collect your activity data but you received the same features and user experience from their products, how much would you be willing to pay for such a service?
    
    \textcolor{note}{\emph{[Slider from \$0 to \$100 per month]}}
    
\end{questions}

\paragraph{Activity Explanations}\mbox{}\\
Google gives different explanations for why they collect activity data. They differentiate between three categories of activities: Web activities, YouTube activities, and Maps activities.
The following shows for each of these categories the explanations Google gives for why they store activity data.

\pagebreak
\begin{questions}[resume]
    \item Do you think this is an appropriate reason to store your Google Search activity?
    \label{app:main-survey:Q17}
    \begin{multicols}{2}
        \begin{answers}
            \item Absolutely inappropriate
            \item Inappropriate
            \item Slightly inappropriate
            \item Neutral
            \item Slightly appropriate
            \item Appropriate
            \item Absolutely appropriate
        \end{answers}
    \end{multicols}
    
    \item Do you think this is an appropriate reason to store your YouTube activity?
    \label{app:main-survey:Q18}
    \begin{multicols}{2}
        \begin{answers}
            \item Absolutely inappropriate
            \item Inappropriate
            \item Slightly inappropriate
            \item Neutral
            \item Slightly appropriate
            \item Appropriate
            \item Absolutely appropriate
        \end{answers}
    \end{multicols}
    
    \item Do you think this is an appropriate reason to store your Google Maps activity?
    \label{app:main-survey:Q19}
    \begin{multicols}{2}
        \begin{answers}
            \item Absolutely inappropriate
            \item Inappropriate
            \item Slightly inappropriate
            \item Neutral
            \item Slightly appropriate
            \item Appropriate
            \item Absolutely appropriate
        \end{answers}
    \end{multicols}
\end{questions}

\paragraph{Auto-Delete Options}\mbox{}\\
Google allows you to change how long your online activity is stored.
These settings are called auto-delete options and can be used to automatically delete activities older than a set amount of time. 

\begin{questions}[resume]
    \item Would you like to change how long your activities are stored?
    \label{app:main-survey:Q20}

    \begin{questions}
        \item Please explain if and why you would like to change how long your activities are stored?
        \begin{answers}[widest=Answer:]
            \item[Answer:] \hrulefill
        \end{answers}
    \end{questions}

    \item Google provides a way for you to pause collection of activity data, what do you believe happens when you pause activity data collection?
    \label{app:main-survey:Q21}
    \begin{answers}
        \item Google no longer collects activity data about me
        \item Google still collects activity data about me, but does not associate it with my account
        \item Google still collects activity data about me and still associates it with my account, but simply does not display it on the My Activity page.
        \item Other: \hrulefill
    \end{answers}
    
    \item[\textbf{A2}] What is the color of a red ball?
    \begin{multicols}{2}
        \begin{answers}
            \item Red
            \item Round
            \item Blue
            \item Square
        \end{answers}
    \end{multicols}
    
    \item Do you think My Activity helps you to better understand what data Google collects about you?
    \label{app:main-survey:Q22}
    \begin{multicols}{2}
        \begin{answers}
            \item Strongly disagree
            \item Disagree
            \item Neither agree nor disagree
            \item Agree
            \item Strongly agree
        \end{answers}
    \end{multicols}
    
    \begin{questions}
        \item Please explain why.
        \label{app:main-survey:Q2121a}
        \begin{answers}[widest=Answer:]
            \item[Answer:] \hrulefill
        \end{answers}
    \end{questions}

    \item After completing this survey, do you see yourself changing any setting in your My Activity page?
    \label{app:main-survey:Q23}
    \begin{multicols}{3}
        \begin{answers}
            \item Yes
            \item No
            \item Unsure
        \end{answers}
    \end{multicols}
    \begin{questions}
        \item Which setting, if any, would you change?
        \label{app:main-survey:Q23a}
        \begin{answers}[widest=Answer:]
            \item[Answer:] \hrulefill
        \end{answers}
    \end{questions}
    
    \item In a month, do you see yourself reviewing and/or deleting activities using My Activity?
    \label{app:main-survey:Q24}
    \begin{multicols}{3}
        \begin{answers}
            \item Yes
            \item No
            \item Unsure
        \end{answers}
    \end{multicols}
    \begin{questions}
        \item Which kinds of activities, if any, would you review and/or delete?
        \label{app:main-survey:Q24a}
        \begin{answers}[widest=Answer:]
            \item[Answer:] \hrulefill
        \end{answers}
    \end{questions}
    
    \item \label{app:main-survey:Q25} Now that you have explored, My Activity, do you plan using Google products differently in the future?
    \begin{multicols}{3}
        \begin{answers}
            \item Yes
            \item No
            \item Unsure
        \end{answers}
    \end{multicols}
    
    \begin{questions}
        \item What would you change and why?
        \label{app:main-survey:Q25a}
        
        \textcolor{note}{\emph{[Shown only if answer to \ref{app:main-survey:Q25} was \enquote{Yes}]}}
        \begin{answers}[widest=Answer:]
            \item[Answer:] \hrulefill
        \end{answers}
        
        \item Why are you unsure if you would change using Google products?
        \label{app:main-survey:Q25b}
        
        \textcolor{note}{\emph{[Shown only if answer to \ref{app:main-survey:Q25} was \enquote{Unsure}]}}
        \begin{answers}[widest=Answer:]
            \item[Answer:] \hrulefill
        \end{answers}
        
        \item Why would you not change using Google products?
        \label{app:main-survey:Q25c}
        
        \textcolor{note}{\emph{[Shown only if answer to \ref{app:main-survey:Q25} was \enquote{No}]}}
        \begin{answers}[widest=Answer:]
            \item[Answer:] \hrulefill
        \end{answers}
    \end{questions}
    
    \item How concerned are you with the amount of information Google is collecting about your activities online?
    \label{app:main-survey:Q26}
    \begin{multicols}{2}
        \begin{answers}
            \item Not at all concerned
            \item Slightly concerned
            \item Somewhat concerned
            \item Moderately concerned
            \item Extremely concerned
        \end{answers}
    \end{multicols}
    \begin{questions}
        \item Please explain why.
        \label{app:main-survey:Q26a}
        \begin{answers}[widest=Answer:]
            \item[Answer:] \hrulefill
        \end{answers}
    \end{questions}
    
    \item How often do you benefit from the amount of information that Google collects about your activities online?
    \label{app:main-survey:Q27}
    \begin{multicols}{2}
        \begin{answers}
            \item Never
            \item Rarely
            \item Sometimes
            \item Often
            \item Always
        \end{answers}
    \end{multicols}
    
    \begin{questions}
        \item Please explain why.
        \label{app:main-survey:Q27a}
        \begin{answers}[widest=Answer:]
            \item[Answer:] \hrulefill
        \end{answers}
    \end{questions}
    
\end{questions}

\begin{questions}[label=\textbf{D\arabic*}]
    \item What is your gender?
    \label{app:main-survey:D1}
    \begin{multicols}{2}
        \begin{answers}
            \item Woman
            \item Man
            \item Non-binary
            \item Prefer not to disclose
            \item Prefer to self-describe
        \end{answers}
    \end{multicols}
    
    \item What is your age?
    \label{app:main-survey:D2}
    \begin{multicols}{2}
        \begin{answers}
            \item 18 -- 24
            \item 25 -- 34
            \item 35 -- 44
            \item 45 -- 54
            \item 55 -- 64
            \item 65 or older
            \item Prefer not to disclose
        \end{answers}
    \end{multicols}
    
    \item What is the highest degree or level of school you have completed?
    \label{app:main-survey:D3}
    \begin{answers}
        \item No schooling completed
        \item Some high school, no diploma
        \item High school graduate, diploma, or equivalent
        \item Some college credit, no degree
        \item Trade / technical / vocational training
        \item Associate degree
        \item Bachelor's degree
        \item Master's degree
        \item Professional degree (\eg, J.D., M.D.)
        \item Doctorate degree
        \item Prefer not to disclose
    \end{answers}

    \item Which of the following best describes your educational background or job field?\label{app:main-survey:D4}
    \begin{answers}
        \item I have an education in, or work in, the field of computer science, computer engineering or IT.
        \item I do not have an education in, nor do I work in, the field of computer science, computer engineering or IT.
        \item Prefer not to disclose
    \end{answers}
\end{questions}

\end{scriptsize}

\clearpage

\subsection{Qualitative Codes}\label{app:codebook}
\begin{itemize}
\item\textbf{ad-blocker (1)}

\item\textbf{against-collection (25)}

\emph{activities, computer-usage, maps, not-necessary, phone-usage, risky, search-history, selling-data, website-history, youtube}

\item\textbf{against-data-collection (5)}

\item\textbf{against-filter-bubble (2)}

\item\textbf{against-targeted-advertising (14)}

\item\textbf{amount (85)}

\emph{as-expected, less-than-expected, more-than-expected}

\item\textbf{artificial-intelligence-concerns (3)}

\emph{biased-algorithms}

\item\textbf{behavior-modification (1)}

\item\textbf{being-used-to-make-money (5)}

\item\textbf{benefit (152)}

\item\textbf{better-understanding (115)}

\emph{google-profit-motive, how-long-information-is-stored, how-much-information-is-collected, how-to-change-settings, implications-of-data-collection, inferences, myactivity-existence, other-google-apps, usage-history, what-google-is-doing, what-information-is-collected, when-information-is-collected, why-information-is-collected}

\item\textbf{browser (38)}

\emph{ad-blocker, block-cookies, clear-history-cache, extension, popup-blocker, private-window, remove-cookies, second-browser, tracker-blocking}

\item\textbf{change-settings (98)}

\emph{all-settings, auto-delete, automaticaly-delete-data, data-storage-length, delete-activities-more-often, delete-frequency, google-my-activity, history, how-long-information-stored, limit-data-collection, maps, never-delete, reminders, search, stop-data-collection, turn-off-collection, turn-on-more-data-collection, youtube}

\item\textbf{change-use (16)}

\emph{more-careful, use-non-google-browser, use-non-google-search, use-non-google-services}

\item\textbf{collection-aware (11)}

\item\textbf{collection-beneficial (67)}

\emph{better-recommendations, easy-to-find-activity, improved-experience, improves-services, maps, personalization, personalized-ads, preferences, recommendations, revisit-activities, search, search-history, shortcuts, youtube}

\item\textbf{collection-not-beneficial (5)}

\item\textbf{collection-unnecessary (2)}

\item\textbf{concern (6)}

\item\textbf{control (1)}

\item\textbf{convenient (22)}

\emph{but-not-enough}

\item\textbf{cookies (1)}

\item\textbf{creepy (4)}

\item\textbf{data-collection (93)}

\emph{interests, internet-use-history, location, user-information}

\item\textbf{data-collection-beneficial (1)}

\item\textbf{delete-activity (90)}

\emph{all, app-data, history, irrelevant, maps, music, search, sensitive, web, youtube}

\item\textbf{delete-data (12)}

\emph{all, banking, history, web-activity, work-related-information}

\item\textbf{delete-immediately (5)}

\emph{web-history}

\item\textbf{delete-information (3)}

\emph{google}

\item\textbf{details (17)}

\emph{more-than-expected}

\item\textbf{did-not-change-perspective (5)}

\item\textbf{dont-store-activities-time-period (40)}

\emph{day-or-two, forever, long, maps, one-month, one-week, short-period, three-months}

\item\textbf{experience-improvements (109)}

\emph{ease-of-use, maps, personalization, saving-history, search, user-interface}

\item\textbf{false-inferences (4)}

\item\textbf{google-monopoly (4)}

\item\textbf{google-products-convenient (5)}

\emph{google-account-login}

\item\textbf{google-products-necessary (18)}

\emph{better-than-alternatives, daily, habit, habits, life, school, tied-to-google, work}

\item\textbf{government (1)}

\item\textbf{happy-with-google (3)}

\item\textbf{happy-with-status-quo (37)}

\item\textbf{harm (1)}

\item\textbf{history (43)}

\emph{useful}

\item\textbf{i-am-the-product (1)}

\item\textbf{increase-google-product-use (2)}

\item\textbf{increased-awareness (3)}

\emph{after-exploring-myactivity, google-data-collection, search}

\item\textbf{interested (9)}

\item\textbf{liability-coverage (1)}

\item\textbf{limit-information (25)}

\emph{amount, gdpr-opt-out, location, sensitive, sharing}

\item\textbf{limit-usage (21)}

\emph{account, accounts, anonymize-activity, gmail, google-products, google-chrome, google-products, google-services, no-chrome, no-click-ads, no-store-information, restrict-search, search, third-party-aps}

\item\textbf{little-benefit (26)}

\item\textbf{make-money (68)}

\emph{sell-data}

\item\textbf{makes-note-of-offending-website (1)}

\item\textbf{modify-settings-or-configuration (17)}

\emph{ad-personalization-off, data-gathering, google, privacy-settings}

\item\textbf{my-activity-helpful (16)}

\item\textbf{myactivity-useful (14)}

\emph{activitiy-history, detailed-information, for-google, metrics, one-location-for-activity-history, recommendations, search}

\item\textbf{never-delete (2)}

\emph{maps, youtube}

\item\textbf{no-answer (1)}

\item\textbf{no-benefit (45)}

\item\textbf{no-change (105)}

\emph{already-configured, likes-current-settings, too-time-consuming, would-forget}

\item\textbf{no-sentiment (40)}

\item\textbf{no-strategies (78)}

\emph{doesnt-know-how, overwhelmed, wants-strategies}

\item\textbf{no-surprise (34)}

\item\textbf{none (1)}

\item\textbf{not-better-understanding (13)}

\emph{already-know-activity}

\item\textbf{not-relevant-to-store (42)}

\emph{a-few-months, eighteen-months, five-years, maps, one-month, one-week, one-year, search, three-months, youtube}

\item\textbf{nothing-to-hide (23)}

\item\textbf{personalized-ads (84)}

\emph{creepy, dislike, manipulation, not-useful, partial-useful, useful}

\item\textbf{physical-safety-concerns (2)}

\emph{location-data}

\item\textbf{privacy-aware (36)}

\emph{amount-of-data-collected}

\item\textbf{privacy-concerns (253)}

\emph{amount-of-information, behavior-modification, biographical-information, control-data, data-collection, data-future, data-sensitive, false-inferences, feels-violated, google-pervasive, government, how-long-data-stored, inferences, information-collection, information-used-against-me, information-used-against-you, invasion-of-privacy, invasive, lack-of-consent, lack-of-control, location, monitoring, not-properly-anonomized-data, search-history-used-against-me, selling-data, selling-information, sensitive-information, stalking, third-parties, tracking, uncomfortable-sharing}

\item\textbf{privacy-protection (1)}

\item\textbf{privacy-resigned (34)}

\emph{change-wont-make-difference, data-collection-unavoidable}

\item\textbf{privacy-tradeoff (19)}

\emph{convenience, free-services}

\item\textbf{product-improvements (42)}

\emph{auto-complete, products-and-services, search}

\item\textbf{profit-form-data (14)}

\emph{get-paid, little-in-return, services-are-free}

\item\textbf{protect-google (1)}

\item\textbf{request-data (1)}

\item\textbf{research (15)}

\emph{machine-learning}

\item\textbf{review (11)}

\emph{google-data-collection-policies, google-myactivity, history, work-related}

\item\textbf{scared (2)}

\item\textbf{search-bubble (3)}

\item\textbf{security (9)}

\emph{accounts, passwords}

\item\textbf{security-concerns (85)}

\emph{data-breach, data-future, data-misuse, data-released, location}

\item\textbf{skeptical (31)}

\emph{designed-to-confuse, does-not-show-all-data-collection, google-only-helping-itself, undisclosed-use-of-data}

\item\textbf{social-influence (1)}

\item\textbf{societal-impact (2)}

\item\textbf{some-benefit (55)}

\item\textbf{strategies (1)}

\emph{fake-info}

\item\textbf{suggestions (149)}

\emph{not-useful, partial-useful, useful}

\item\textbf{suggestions-not-accurate (1)}

\emph{maps, youtube}

\item\textbf{surprise (51)}

\item\textbf{targeted-advertising (123)}

\item\textbf{too-much-data (2)}

\item\textbf{too-much-time-required-to-change (5)}

\item\textbf{tools (17)}

\emph{antimalware, google-activity-control, google-privacy-checkup, non-tracking-search-engine, secure-server, vpn}

\item\textbf{trust-google (33)}

\item\textbf{uncertain (12)}

\emph{about-benefit, how-much-is-collected, what-is-collected}

\item\textbf{unclear (8)}

\item\textbf{unclear-statement (15)}

\item\textbf{uncomfortable (4)}

\item\textbf{unconcerned (131)}

\emph{collection, have-control, my-activity-not-a-risk, non-sensitive}

\item\textbf{undecided (57)}

\emph{how-much-work-to-change, if-necessary, needs-more-review, privacy-level, wants-to-review-data-collection, wants-to-review-devices, wants-to-review-history, wants-to-review-settings, what-information-to-save, what-settings-to-change}

\item\textbf{undisclosed-collection (3)}

\item\textbf{unknown (7)}

\item\textbf{unknowns (98)}

\emph{amount-data-collected, how-information-collected, how-information-used, how-long-information-stored, how-much-information, how-often-data-collected, how-secure-is-information, how-to-control, how-to-protect-privacy, inferences, information-collected, information-future, risk-versus-benefit, what-data-collected, what-information-collected, when-information-collected, where-information-collected, where-information-goes, who-has-access, why-information-collected, why-information-stored, why-myactivity, why-recent-web-search-not-shown}

\item\textbf{unsure (1)}

\item\textbf{use-false-information (6)}

\emph{alias, fake-account}

\item\textbf{wants-control-over-what-is-stored (4)}

\end{itemize}

\clearpage

\onecolumn
\subsection{Demographics}\label{app:demographics}
\begin{table*}[ht]
  \centering
\caption[Full demographics]{\label{tab:demographics-full}
Full demographic data of the participants of the main survey.
}
\small
\renewcommand{\arraystretch}{0.6}
\begin{tabular*}{\textwidth}{
>{\bfseries}l
@{\extracolsep{\fill}}
r
*{8}{r}
}
\toprule
& & \multicolumn{2}{c}{\textbf{Male}} 
  & \multicolumn{2}{c}{\textbf{Female}} 
  & \multicolumn{2}{c}{\textbf{Other}} 
  & \multicolumn{2}{c}{\textbf{Total}} \\    
 \cmidrule(lr){3-4} \cmidrule(lr){5-6} \cmidrule(lr){7-8} \cmidrule(lr){9-10}
& & \multicolumn{1}{c}{\textbf{n}} & \multicolumn{1}{c}{\textbf{\%}} 
  & \multicolumn{1}{c}{\textbf{n}} & \multicolumn{1}{c}{\textbf{\%}} 
  & \multicolumn{1}{c}{\textbf{n}} & \multicolumn{1}{c}{\textbf{\%}} 
  & \multicolumn{1}{c}{\textbf{n}} & \multicolumn{1}{c}{\textbf{\%}} \\
\midrule
\multirow{7.25}{*}{Age}
& 18--24 & 14     & 9    & 13    & 8    & 2 & 1&   29 & 19\\
& 25--34 & 20     & 13    & 15    & 10    & 0 & 0 &   35 & 23 \\
& 35--44 & 17     & 11    & 14    & 9     & 0 & 0 &   31 & 20 \\
& 45--54 & 14     & 9     & 15    & 10    & 0 & 0 &   29 & 19 \\
& 55--64 & 12     & 8     & 12    & 8     & 0 & 0 &   24 & 16  \\
& 65 or older & 2      & 1     & 2     & 1     & 0 & 0 &   4  & 3  \\
& No answer   & 0 & 0 & 0 & 0 & 1 & 1 & 1 & 1  \\
\midrule
\multirow{10.25}{*}{Education}
& Some High School    & 1   & 1   & 0     & 0     & 0 & 0 &    1 & 1 \\
& High School         & 12  & 8   & 2    & 1     & 1 & 1 &   15 & 10 \\
& Some College        & 15 & 10  & 15   & 10    & 1 & 1 &   31 & 20 \\
& Training            & 1  & 1   & 1    & 1     & 0 & 0 &   2 & 1 \\
& Associates          & 7  & 5   & 8    & 5     & 0 & 0 &   15 & 10  \\
& Bachelor's          & 28 & 18  & 27   & 18    & 0 & 0 &   55 & 36  \\
& Master's            & 11  & 7   & 13    & 8     & 0 & 0 &   24 & 16  \\
& Professional        & 1  & 1   & 1     & 1     & 0 & 0 &   2 & 1  \\
& Doctorate           & 3  & 2   & 4     & 3     & 0 & 0 &   7 & 5  \\
& No answer   & 0   & 0   & 0     & 0     & 1 & 1 &   1 & 1  \\
\midrule
\multirow{3.25}{*}{\parbox[c]{35pt}{\centering Back\-ground}}
& Tech                     & 29 & 19  &  8 &  5  & 0 & 0 &  37 & 24 \\
& No Tech                  & 48 & 31  & 59 & 39  & 2 & 1 & 109 & 71 \\
& No answer   &  2 &  1  &  4 &  3  & 1 & 1 &  7  &  5 \\
\midrule
& \textbf{Total} & 79 & 52 & 71 & 46 & 3 & 2 & 153 & 100 \\
\bottomrule
\end{tabular*}
\end{table*}

\clearpage

\subsection{Regression Tables}\label{app:regression-tables}
\setlength{\textfloatsep}{2pt plus 1.0pt minus 2.0pt}

\begin{table*}[ht]
\centering
\caption[Post-exposure concern regression analysis]{\label{tab:post-concern-regression-full}
Ordinal regression model to describe the level of concern after visiting the My Activity dashboard.
The model uses a descending concern scale (\ie, from \emph{extremely} to \emph{not at all concerned}).
The Aldrich-Nelson pseudo $R^2$ of the model is \num[round-mode=places]{0.631858309728864}.
}
{
\small
\renewcommand{\arraystretch}{0.6}
\begin{tabular*}{\textwidth}{
l
@{\extracolsep{\fill}}
S[table-format = -1.2, parse-numbers = false, parse-units = false]
S[table-format = 3.2, scientific-notation = fixed, round-mode = places, round-minimum = 0.01, round-integer-to-decimal, tight-spacing = true]
S[table-format = 1.2, parse-numbers = false, parse-units = false]
S[table-format = -1.2, parse-numbers = false, parse-units = false]
S[table-format = <1.3, scientific-notation = fixed, round-mode = places, round-precision = 3, round-minimum = 0.001]
@{\extracolsep{6pt}}
l
}
    \toprule
    {\textbf{Factor}} & {\textbf{Estimate}} & {\textbf{Odds ratio}} & {\textbf{Error}} & {\textbf{t value}} & {\textbf{Pr(\textgreater\textbar z\textbar)}} & \\ 
    \midrule
    $\text{Pre-Exposure concern} = \textit{Extremely}$ & 5.71 & 302.8410699 & 1.03 & 5.56 & 2.74E-08 & *** \\ 
    $\text{Pre-Exposure concern} = \textit{Moderately}$ & 4.56 &  95.6011912 & 0.94 & 4.87 & 1.09E-06 & *** \\ 
    $\text{Pre-Exposure concern} = \textit{Somewhat}$ & 2.77 &  15.9609651 & 0.88 & 3.14 & 1.69E-03 & ** \\ 
    $\text{Pre-Exposure concern} = \textit{Slightly}$ & 1.18 &   3.2585878 & 0.86 & 1.38 & 1.67E-01 &  \\ 
    Increasing benefit & -0.17 &   0.8475816 & 0.37 & -0.45 & 6.54E-01 &  \\ 
    $\text{Knows My Activity} = Yes$ & -0.32 &   0.7247347 & 0.34 & -0.94 & 3.48E-01 &  \\ 
    $\text{IUIPC control} > 3.5$ & 0.29 &   1.3422555 & 1.07 & 0.27 & 7.84E-01 &  \\ 
    $\text{IUIPC awareness} > 3.5$ & -0.41 &   0.6607805 & 2.07 & -0.20 & 8.42E-01 &  \\ 
    $\text{IUIPC collection} > 3.5$ & 0.29 &   1.3317022 & 0.54 & 0.53 & 5.95E-01 &  \\ 
    $\text{Gender} = Male$ & -0.24 &   0.7865547 & 0.34 & -0.71 & 4.81E-01 &  \\ 
    $\text{Age} \in\{18-34,~25-34\}$ & 0.39 &   1.4707258 & 0.48 & 0.81 & 4.17E-01 &  \\ 
    $\text{Age} \in\{35-44,~45-54\}$ & 0.91 &   2.4956446 & 0.47 & 1.95 & 5.15E-02 & $\cdot$ \\ 
    $\text{Education} \in \{\textit{No schooling, (some) high school}\}$ & 0.19 &   1.2054607 & 0.55 & 0.34 & 7.34E-01 &  \\ 
    $\text{Education} \in \{\textit{Some college, Associate, Professional}\}$ & 0.07 &   1.0735471 & 0.36 & 0.20 & 8.44E-01 &  \\ 
    $\text{Has IT background}$ & 0.64 &   1.9011462 & 0.40 & 1.62 & 1.05E-01 &  \\ 
    $\text{\# of activities} > \text{median}$ & -0.47 &   0.6269355 & 0.32 & -1.45 & 1.46E-01 &  \\ 
    \midrule
    \textbf{Intercepts} & & & & & \\
    \midrule
    $\textit{Not at all concerned}~|~\textit{Slightly concerned}$ & 0.58 &   1.7861953 & 2.01 & 0.29 & 7.73E-01 &  \\ 
    $\textit{Slightly concerned}~|~\textit{Somewhat concerned}$ & 2.87 &  17.6636336 & 2.03 & 1.41 & 1.58E-01 &  \\ 
    $\textit{Somewhat concerned}~|~\textit{Moderately concerned}$ & 4.63 & 102.0066655 & 2.05 & 2.26 & 2.40E-02 & * \\ 
    $\textit{Moderately concerned}~|~\textit{Extremely concerned}$ & 6.77 & 875.0841955 & 2.07 & 3.27 & 1.09E-03 & ** \\ 
    \bottomrule
\end{tabular*}
}
\footnotesize
\textbf{Signif. codes:} $\text{***}~\widehat{=} < 0.001$; $\text{**}~\widehat{=} <0.01$; $\text{*}~\widehat{=} <0.05$; $\cdot~\widehat{=} <0.1$
\end{table*}

\begin{table*}[t]
\centering
\caption[Post-exposure frequency of benefit regression analysis]{\label{tab:post-benefit-regression-full}
Ordinal regression model to describe the frequency of benefit after after visiting the My Activity dashboard.
In the model a ascending frequency scale (\ie, from \emph{never} to \emph{always}) is used.
The Aldrich-Nelson pseudo $R^2$ of the model is \num[round-mode=places]{0.681841649836333}.
}
{
\small
\renewcommand{\arraystretch}{0.75}
\begin{tabular*}{\textwidth}{
l
@{\extracolsep{\fill}}
S[table-format = -1.2, parse-numbers = false, parse-units = false]
S[table-format = 1.2e+1, tight-spacing = true]
S[table-format = 1.2, parse-numbers = false, parse-units = false]
S[table-format = -1.2, parse-numbers = false, parse-units = false]
S[table-format = <1.3, scientific-notation = fixed, round-mode = places, round-precision = 3, round-minimum = 0.001]
@{\extracolsep{6pt}}
l
}
    \toprule
    {\textbf{Factor}} & {\textbf{Estimate}} & {\textbf{Odds ratio}} & {\textbf{Error}} & {\textbf{t value}} & {\textbf{Pr(\textgreater\textbar z\textbar)}} & \\ 
    \midrule
    $\text{Pre-Exposure benefit} = \textit{Never}$ & 22.29 & 4.81E+09 & 0.70 & 31.72 & 7.42E-221 & *** \\ 
    $\text{Pre-Exposure benefit} = \textit{Rarely}$ & 20.26 & 6.28E+08 & 0.51 & 39.44 & 2.22E-308 & *** \\ 
    $\text{Pre-Exposure benefit} = \textit{Sometimes}$ & 18.58 & 1.17E+08 & 0.52 & 35.95 & 5.37E-283 & *** \\ 
    $\text{Pre-Exposure benefit} = \textit{Often}$ & 16.12 & 1.00E+07 & 0.66 & 24.45 & 4.49E-132 & *** \\ 
    Increasing concern & 0.57 & 1.77E+00 & 0.51 & 1.11 & 2.68E-01 &  \\ 
    $\text{Knows My Activity} = Yes$ & -0.56 & 5.71E-01 & 0.37 & -1.50 & 1.33E-01 &  \\ 
    $\text{IUIPC control} > 3.5$ & 0.00 & 1.00E+00 & 1.02 & 0.00 & 9.98E-01 &  \\ 
    $\text{IUIPC awareness} > 3.5$ & 0.18 & 1.20E+00 & 2.24 & 0.08 & 9.35E-01 &  \\ 
    $\text{IUIPC collection} > 3.5$ & 0.41 & 1.50E+00 & 0.57 & 0.72 & 4.73E-01 &  \\ 
    $\text{Gender} = Male$ & 0.92 & 2.51E+00 & 0.37 & 2.46 & 1.38E-02 & * \\ 
    $\text{Age} \in\{18-34,~25-34\}$ & 0.76 & 2.14E+00 & 0.50 & 1.53 & 1.26E-01 &  \\ 
    $\text{Age} \in\{35-44,~45-54\}$ & 0.58 & 1.79E+00 & 0.51 & 1.13 & 2.56E-01 &  \\ 
    $\text{Education} \in \{\textit{No schooling, (some) high school}\}$ & -0.16 & 8.50E-01 & 0.59 & -0.28 & 7.82E-01 &  \\ 
    $\text{Education} \in \{\textit{Some college, Associate, Professional}\}$ & -0.16 & 8.51E-01 & 0.38 & -0.42 & 6.71E-01 &  \\ 
    $\text{Has IT background}$ & -0.07 & 9.30E-01 & 0.42 & -0.17 & 8.64E-01 &  \\ 
    $\text{\# of activities} > \text{median}$ & 0.30 & 1.36E+00 & 0.34 & 0.89 & 3.73E-01 &  \\ 
    \midrule
    \textbf{Intercepts} & & & & & \\
    \midrule
    $\textit{Always}~|~\textit{Often}$ & 15.61 & 6.00E+06 & 1.68 & 9.28 & 1.69E-20 & *** \\ 
    $\textit{Often}~|~\textit{Sometimes}$ & 18.78 & 1.43E+08 & 1.63 & 11.52 & 1.05E-30 & *** \\ 
    $\textit{Sometimes}~|~\textit{Rarely}$ & 21.99 & 3.55E+09 & 1.66 & 13.27 & 3.32E-40 & *** \\ 
    $\textit{Rarely}~|~\textit{Never}$ & 24.56 & 4.63E+10 & 1.71 & 14.34 & 1.15E-46 & *** \\ 
    \bottomrule
\end{tabular*}
}
\footnotesize
\textbf{Signif. codes:} $\text{***}~\widehat{=} < 0.001$; $\text{**}~\widehat{=} <0.01$; $\text{*}~\widehat{=} <0.05$; $\cdot~\widehat{=} <0.1$

\end{table*}

\begin{table*}[b]
\centering
\caption[changingSettings regression analysis]{\label{tab:change-setting-regression-full}
Binomial logistic model to describe which other factors (beside visiting My Activity) influenced the participants plan to change any Google account settings (\emph{yes} responses to Question \ref{app:main-survey:Q23}).
The Aldrich-Nelson pseudo $R^2$ of the model is \num[round-mode=places]{0.187524841593501}.
}
{
\small
\renewcommand{\arraystretch}{0.75}
\begin{tabular*}{\textwidth}{
l
@{\extracolsep{\fill}}
S[table-format = -2.2, parse-numbers = false, parse-units = false]
S[table-format = 1.2e+1, tight-spacing = true]
S[table-format = 3.2, parse-numbers = false, parse-units = false]
S[table-format = -1.2, parse-numbers = false, parse-units = false]
S[table-format = <1.3, scientific-notation = fixed, round-mode = places, round-precision = 3, round-minimum = 0.001]
@{\extracolsep{6pt}}
l
}
    \toprule
    {\textbf{Factor}} & {\textbf{Estimate}} & {\textbf{Odds ratio}} & {\textbf{Error}} & {\textbf{z value}} & {\textbf{Pr(\textgreater\textbar z\textbar)}} & \\ 
    \midrule
    (Intercept) & -14.53 & 4.88E-07 & 882.74 & -0.02 & 9.87E-01 &  \\ 
    Increasing concern & 0.54 & 1.71E+00 & 0.48 & 1.12 & 2.62E-01 &  \\ 
    Increasing benefit & -0.32 & 7.29E-01 & 0.41 & -0.78 & 4.36E-01 &  \\ 
    $\text{IUIPC control} > 3.5$ & -0.32 & 7.27E-01 & 1.02 & -0.31 & 7.55E-01 &  \\ 
    $\text{IUIPC awareness} > 3.5$ & 13.63 & 8.29E+05 & 882.74 & 0.02 & 9.88E-01 &  \\ 
    $\text{IUIPC collection} > 3.5$ & 0.77 & 2.17E+00 & 0.60 & 1.29 & 1.97E-01 &  \\ 
    $\text{Gender} = Male$ & 0.44 & 1.56E+00 & 0.38 & 1.17 & 2.43E-01 &  \\ 
    $\text{Age} \in\{18-34,~25-34\}$ & -0.11 & 8.97E-01 & 0.51 & -0.21 & 8.31E-01 &  \\ 
    $\text{Age} \in\{35-44,~45-54\}$ & 0.09 & 1.10E+00 & 0.51 & 0.18 & 8.55E-01 &  \\ 
    $\text{Education} \in \{\textit{No schooling, (some) high school}\}$ & -0.07 & 9.30E-01 & 0.62 & -0.12 & 9.07E-01 &  \\ 
    $\text{Education} \in \{\textit{Some college, Associate, Professional}\}$ & -0.39 & 6.76E-01 & 0.40 & -0.98 & 3.29E-01 &  \\ 
    $\text{Has IT background}$ & 0.23 & 1.26E+00 & 0.43 & 0.54 & 5.87E-01 &  \\ 
    $\text{\# of activities} > \text{median}$ & -0.27 & 7.66E-01 & 0.35 & -0.76 & 4.50E-01 &  \\ 
    \bottomrule
\end{tabular*}
}
\footnotesize
\textbf{Signif. codes:} $\text{***}~\widehat{=} < 0.001$; $\text{**}~\widehat{=} <0.01$; $\text{*}~\widehat{=} <0.05$; $\cdot~\widehat{=} <0.1$
\end{table*}

\begin{table*}[t]
\centering
\caption[Use Google differently regression analysis]{\label{tab:google-different-regression-full}
Binomial logistic model to describe which other factors (beside visiting My Activity) influenced the participants plan to use Google products differently in the future (\emph{yes} responses to Question \ref{app:main-survey:Q25}).
The Aldrich-Nelson pseudo $R^2$ of the model is \num[round-mode=places]{0.420131390603509}.
}
{
\small
\renewcommand{\arraystretch}{0.75}
\begin{tabular*}{\textwidth}{
l
@{\extracolsep{\fill}}
S[table-format = -2.2, parse-numbers = false, parse-units = false]
S[table-format = 1.2e+1, tight-spacing = true]
S[table-format = 4.2, parse-numbers = false, parse-units = false]
S[table-format = -1.2, parse-numbers = false, parse-units = false]
S[table-format = <1.3, scientific-notation = fixed, round-mode = places, round-precision = 3, round-minimum = 0.001]
@{\extracolsep{6pt}}
l
}   
    \toprule
    {\textbf{Factor}} & {\textbf{Estimate}} & {\textbf{Odds ratio}} & {\textbf{Error}} & {\textbf{z value}} & {\textbf{Pr(\textgreater\textbar z\textbar)}} & \\ 
    \midrule
    (Intercept) & -16.23 & 8.93E-08 & 3956.18 & -0.00 & 9.97E-01 &  \\ 
    Increasing concern & 2.50 & 1.22E+01 & 0.62 & 4.01 & 6.04E-05 & *** \\ 
    Increasing benefit & -1.31 & 2.71E-01 & 0.59 & -2.21 & 2.73E-02 & * \\ 
    $\text{IUIPC control} > 3.5$ & 18.42 & 1.00E+08 & 1361.72 & 0.01 & 9.89E-01 &  \\ 
    $\text{IUIPC awareness} > 3.5$ & -4.82 & 8.06E-03 & 4183.97 & -0.00 & 9.99E-01 &  \\ 
    $\text{IUIPC collection} > 3.5$ & 1.40 & 4.07E+00 & 0.88 & 1.60 & 1.09E-01 &  \\ 
    $\text{Gender} = Male$ & -0.73 & 4.80E-01 & 0.51 & -1.45 & 1.46E-01 &  \\ 
    $\text{Age} \in\{18-34,~25-34\}$ & 0.97 & 2.63E+00 & 0.75 & 1.29 & 1.98E-01 &  \\ 
    $\text{Age} \in\{35-44,~45-54\}$ & 1.21 & 3.36E+00 & 0.76 & 1.60 & 1.09E-01 &  \\ 
    $\text{Education} \in \{\textit{No schooling, (some) high school}\}$ & 0.45 & 1.57E+00 & 0.78 & 0.58 & 5.62E-01 &  \\ 
    $\text{Education} \in \{\textit{Some college, Associate, Professional}\}$ & -0.16 & 8.56E-01 & 0.50 & -0.31 & 7.57E-01 &  \\ 
    $\text{Has IT background}$ & 0.86 & 2.37E+00 & 0.55 & 1.58 & 1.14E-01 &  \\ 
    $\text{\# of activities} > \text{median}$ & -1.25 & 2.86E-01 & 0.46 & -2.72 & 6.49E-03 & ** \\ 
    \bottomrule
\end{tabular*}
}
\footnotesize
\textbf{Signif. codes:} $\text{***}~\widehat{=} < 0.001$; $\text{**}~\widehat{=} <0.01$; $\text{*}~\widehat{=} <0.05$; $\cdot~\widehat{=} <0.1$
\end{table*}

\begin{table*}[tbp]
\centering
\caption[Review activities regression analysis]{\label{tab:review-activity-regression-full}
Binomial logistic model to describe which other factors (beside visiting My Activity) influenced the participants plan to review/delete activities (\emph{yes} responses to Question \ref{app:main-survey:Q24}).
We use the Aldrich-Nelson pseudo $R^2$ as goodness of fit estimate ($R^2 = \num[round-mode=places]{0.25321311230276}$) of the model.
}
{
\small
\renewcommand{\arraystretch}{0.75}
\begin{tabular*}{\textwidth}{
l
@{\extracolsep{\fill}}
S[table-format = -2.2, parse-numbers = false, parse-units = false]
S[table-format = 1.2e+1, tight-spacing = true]
S[table-format = 3.2, parse-numbers = false, parse-units = false]
S[table-format = -1.2, parse-numbers = false, parse-units = false]
S[table-format = <1.3, scientific-notation = fixed, round-mode = places, round-precision = 3, round-minimum = 0.001]
@{\extracolsep{6pt}}
l
}
    \toprule
    {\textbf{Factor}} & {\textbf{Estimate}} & {\textbf{Odds ratio}} & {\textbf{Error}} & {\textbf{z value}} & {\textbf{Pr(\textgreater\textbar z\textbar)}} & \\ 
    \midrule
    (Intercept) & 13.77 & 9.53E+05 & 882.74 & 0.02 & 9.88E-01 &  \\ 
    Increasing concern & 0.90 & 2.45E+00 & 0.49 & 1.82 & 6.80E-02 & . \\ 
    Increasing benefit & 0.05 & 1.05E+00 & 0.41 & 0.13 & 9.00E-01 &  \\ 
    $\text{IUIPC control} > 3.5$ & 0.63 & 1.88E+00 & 1.25 & 0.51 & 6.13E-01 &  \\ 
    $\text{IUIPC awareness} > 3.5$ & -16.69 & 5.63E-08 & 882.74 & -0.02 & 9.85E-01 &  \\ 
    $\text{IUIPC collection} > 3.5$ & 1.43 & 4.19E+00 & 0.71 & 2.03 & 4.24E-02 & * \\ 
    $\text{Gender} = Male$ & 0.45 & 1.56E+00 & 0.39 & 1.14 & 2.55E-01 &  \\ 
    $\text{Age} \in\{18-34,~25-34\}$ & -0.07 & 9.32E-01 & 0.53 & -0.13 & 8.93E-01 &  \\ 
    $\text{Age} \in\{35-44,~45-54\}$ & 0.24 & 1.27E+00 & 0.53 & 0.45 & 6.52E-01 &  \\ 
    $\text{Education} \in \{\textit{No schooling, (some) high school}\}$ & 0.85 & 2.34E+00 & 0.63 & 1.34 & 1.79E-01 &  \\ 
    $\text{Education} \in \{\textit{Some college, Associate, Professional}\}$ & -0.04 & 9.64E-01 & 0.41 & -0.09 & 9.28E-01 &  \\ 
    $\text{Has IT background}$ & 0.34 & 1.40E+00 & 0.44 & 0.77 & 4.41E-01 &  \\ 
    $\text{\# of activities} > \text{median}$ & -0.35 & 7.02E-01 & 0.36 & -0.97 & 3.30E-01 &  \\ 
    \bottomrule
\end{tabular*}
}
\footnotesize
\textbf{Signif. codes:} $\text{***}~\widehat{=} < 0.001$; $\text{**}~\widehat{=} <0.01$; $\text{*}~\widehat{=} <0.05$; $\cdot~\widehat{=} <0.1$
\end{table*}

\clearpage

\subsection{Screenshots of My Activity and the Survey}\label{app:screenshots}

\begin{figure*}[ht]
\centering
\includegraphics[width=\columnwidth]{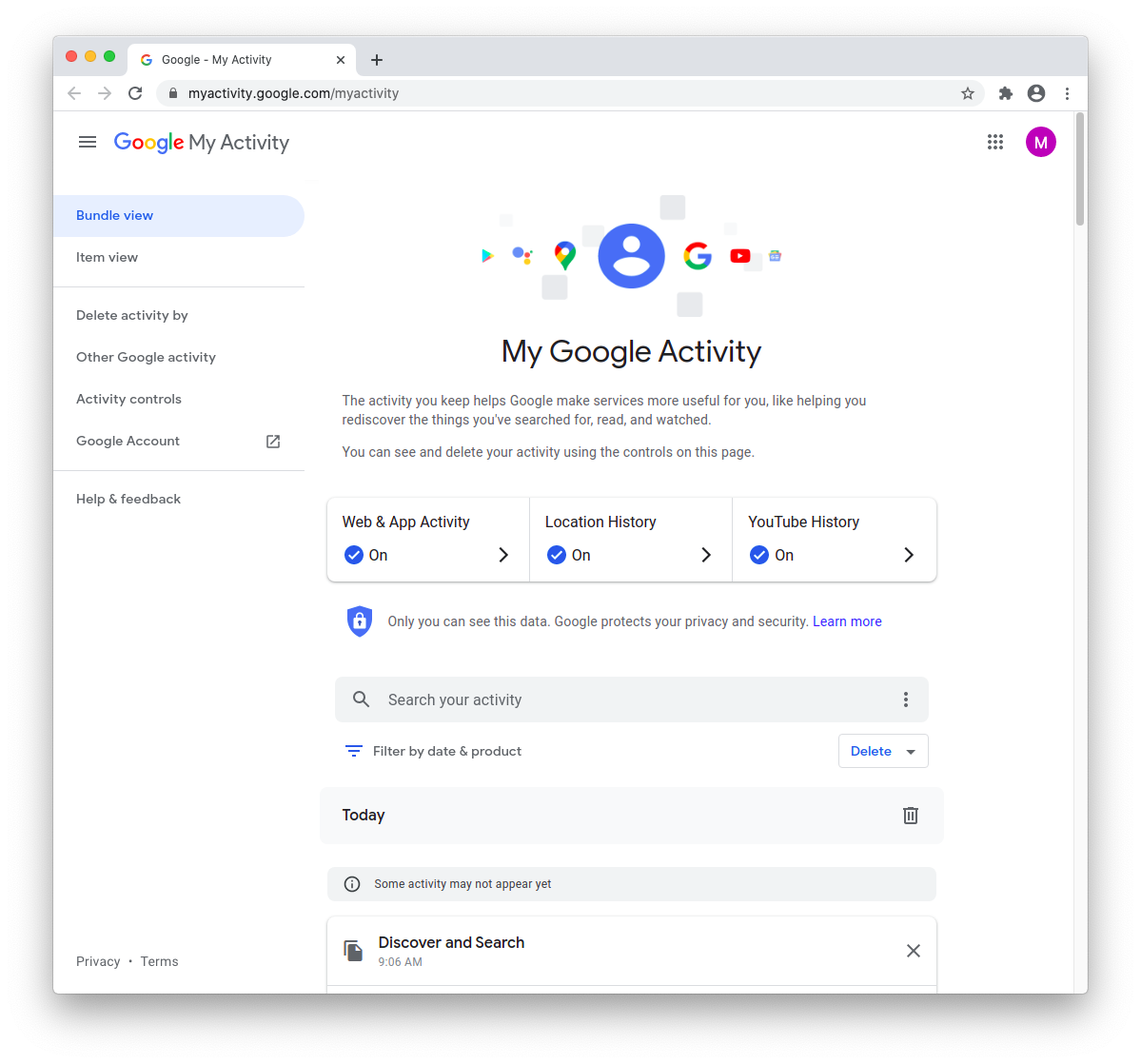}
\caption[My Activity user interface]{\label{app:fig:my-activity}
The My Activity user interface at the time of the study in September 2020.
}
\end{figure*}

\begin{figure*}[ht]
\centering
\includegraphics[width=\columnwidth]{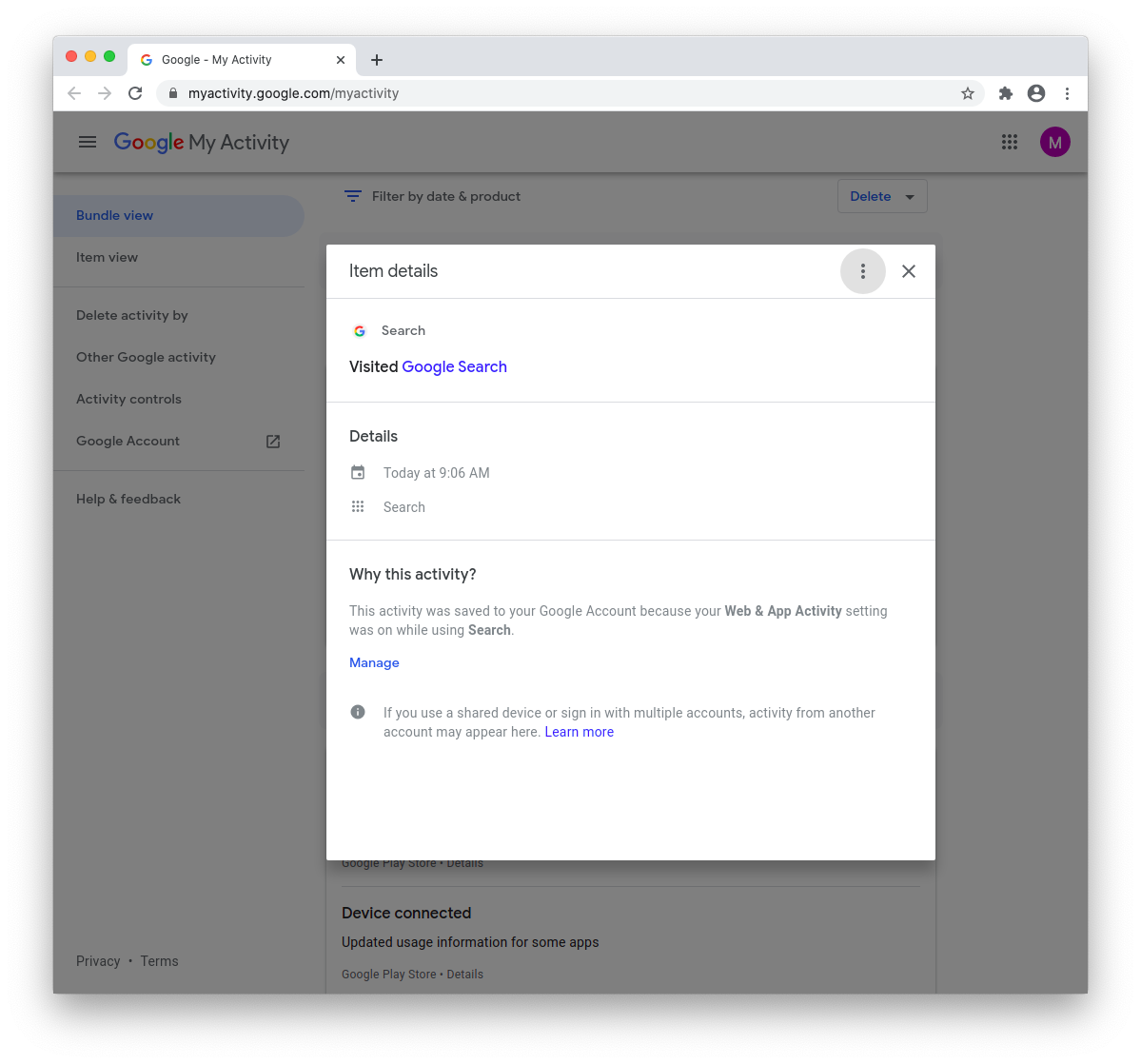}
\caption[My Activity item details]{\label{app:fig:item-details}
Clicking on the details link of an activity opens an additional view providing information like the time and date, type of the activity, and in which service or app the activity was collected.
}
\end{figure*}

\begin{figure*}[ht]
\centering
\includegraphics[width=\columnwidth]{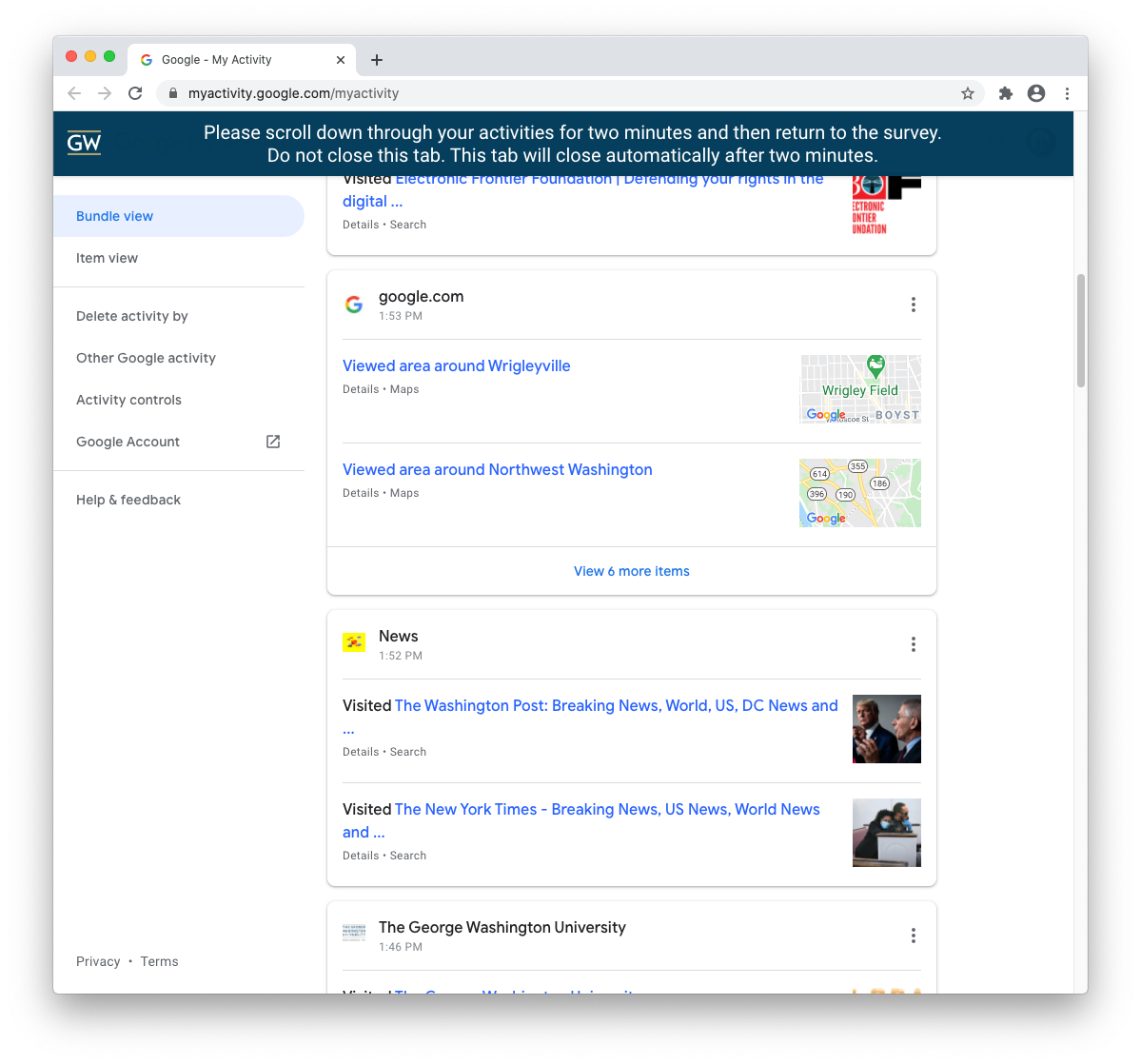}
\caption[Exploring My Activity during the survey]{\label{app:fig:explor-my-activity}
During the My Activity exploration phase of the survey, we added a banner at the top of the My Activity page and disabled all buttons and hyperlinks on the page. 
}
\end{figure*}

\begin{figure*}[ht]
\centering
\includegraphics[width=\columnwidth]{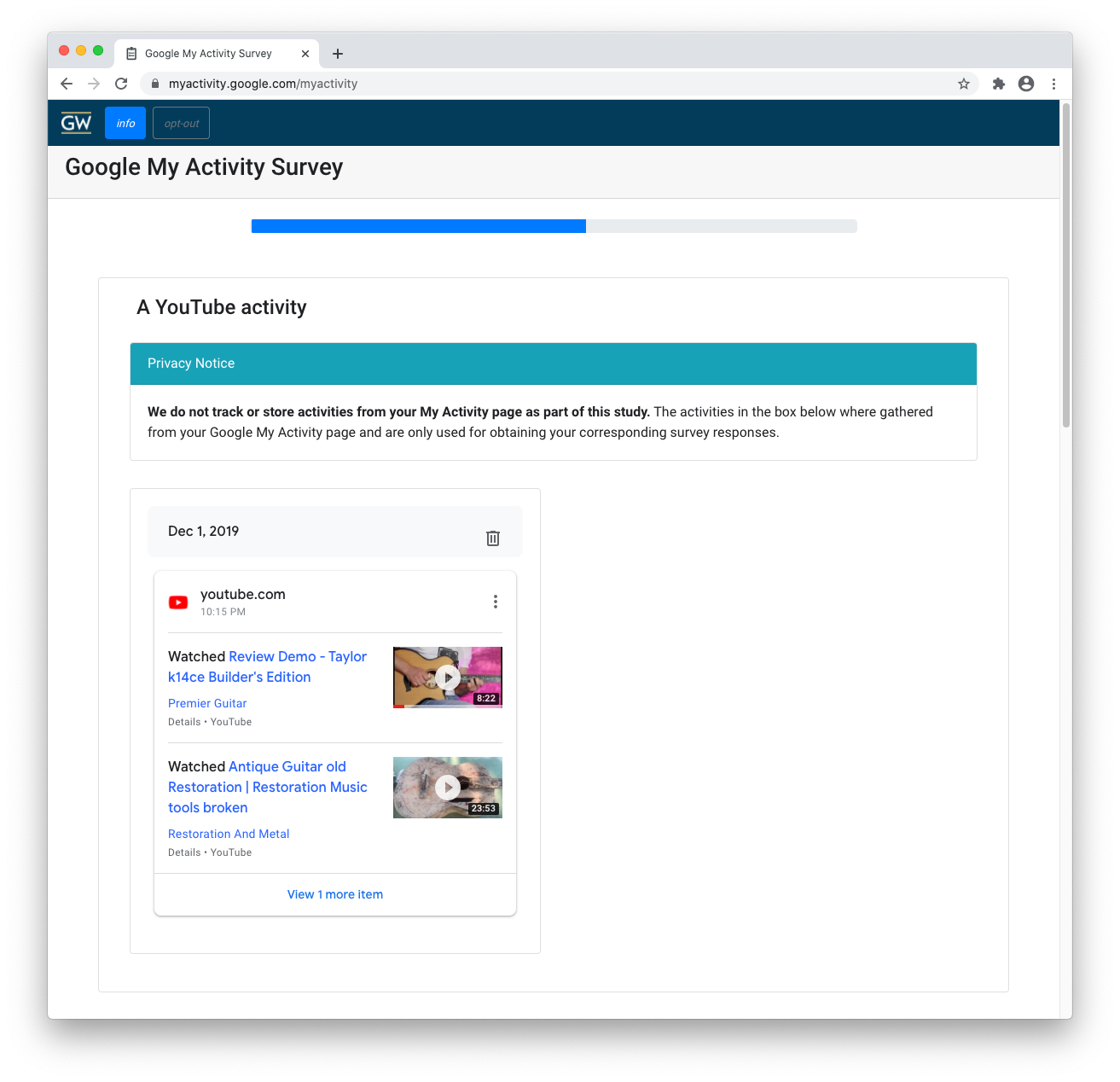}
\caption[Activity presentation in the survey]{\label{app:fig:survey-activities}
In the Activity Presentation phase of the survey, we embedded activity bundles into the survey page. The participants saw this type of page up to 9 times. Every time with another activity bundle.
}
\end{figure*}

\clearpage

\end{document}